\documentclass[twocolumn,floatfix,superscriptaddress,amsmath,showpacs,showkeys,aps,prb]{revtex4}

\usepackage[utf8]{inputenc}
\usepackage[final]{graphicx}
\usepackage{t1enc}
\usepackage{bm}

\begin{document}

\title{\bf Interplay between coarsening and nucleation in an Ising model
with dipolar interactions}

\author{Sergio A. Cannas}
\email{cannas@famaf.unc.edu.ar}
\affiliation{Instituto de Física de la Facultad de Matemática, Astronomía y Física (IFFAMAF-CONICET),
Universidad Nacional de Córdoba \\
Ciudad Universitaria, 5000 Córdoba, Argentina}
\author{Mateus F. Michelon}
\email{michelon@ifi.unicamp.br}
\affiliation{Institituo de Física Gleb Wataghin, Universidade Estadual
de Campinas\\
CP 6165, 13083-970 Campinas, SP, Brazil}
\author{Daniel A. Stariolo}
\email{stariolo@if.ufrgs.br}
\affiliation{Departamento de Física,
Universidade Federal do Rio Grande do Sul\\
CP 15051, 91501-970 Porto Alegre, RS, Brazil}
\altaffiliation{Research Associate of the Abdus Salam International Centre for
Theoretical Physics, Trieste, Italy}
\author{Francisco A. Tamarit}
\email{tamarit@famaf.unc.edu.ar}
\affiliation{Instituto de Física de la Facultad de Matemática, Astronomía y Física (IFFAMAF-CONICET),
Universidad Nacional de Córdoba\\
Ciudad Universitaria, 5000 Córdoba, Argentina}

\date{\today}

\begin{abstract}
We study the dynamical behavior of a square lattice Ising model with exchange
and dipolar
interactions by means of Monte Carlo simulations. After a sudden quench to low
temperatures we find that the system may undergo a coarsening process where
stripe phases with different orientations compete or alternatively it can relax
initially to a metastable nematic phase and then decay to the equilibrium stripe
phase through nucleation. We measure the distribution of equilibration times for
both processes and compute their relative probability of occurrence as a
function of temperature and system size. This peculiar relaxation mechanism is
due to the strong metastability of the
nematic phase, which goes deep in the low temperature stripe phase. We also measure 
quasi-equilibrium
autocorrelations in a wide range of temperatures. They show a distinct decay to
a plateau that we identify as due to a finite fraction of frozen spins in the
nematic phase. We find  indications that the plateau is a finite size effect.
Relaxation times as a function of temperature in the metastable region show super-Arrhenius behavior,
suggesting a possible glassy behavior of the system at low temperatures.
\end{abstract}

\pacs{64.60.qe,64.75.St,75.40.Gb,75.40.Mg}
\keywords{dipolar interactions,  Ising model, nucleation, coarsening}

\maketitle

\section{Introduction}

It is well known that the usual long range order characteristic of
a ferromagnetic phase dominated by exchange interactions can be destroyed by the
presence of dipolar interactions inducing the formation of magnetic
domains~\cite{HuSc1998}. In some cases, typically when perpendicular anisotropy
is present, the dipolar interactions are antiferromagnetic in
character and compete with the ferromagnetic exchange interactions
~\cite{Po1998,GiLeLi2006} giving rise
to antiferromagnetic or stripe phases. These phases are common in ferromagnetic
thin films with perpendicular anisotropy and have been the subject of
intense experimental~\cite{AlStBi1990,
VaStMaPiPoPe2000,WuWoSc2004,WoWuCh2005,PoVaPe2006}, theoretical~\cite{YaGy1988,
PePo1990,AbKaPoSa1995} and numerical~\cite{MaWhRoDe1995,CaStTa2004,CaMiStTa2006,
RaReTa2006,PiCa2007,CaBiPiCaStTa2008} work in the last 20 years. In the strong anisotropy limit
an ultrathin film can be modeled by a system of magnetic dipoles on a square
lattice in which the magnetic moments
 are oriented perpendicular to the plane of the lattice, with both nearest-neighbor
ferromagnetic exchange interactions and long-range dipole-dipole
interactions between moments. The thermodynamics of this system is
ruled by the dimensionless Ising Hamiltonian:
\begin{equation}
{\cal H}= - \delta \sum_{<i,j>} \sigma_i \sigma_j + \sum_{(i,j)}
\frac{\sigma_i \sigma_j}{r^3_{ij}} \label{Hamilton1}
\end{equation}
\noindent where $\delta$ stands for the ratio between the exchange
$J_0>0$ and the dipolar $J_d>0$ interactions parameters, i.e.,
$\delta = J_0/J_d$. The first sum runs over all pairs of nearest
neighbor spins and the second one over all distinct pairs of spins
of the lattice; $r_{ij}$ is the distance, measured in crystal
units, between sites $i$ and $j$. The energy is measured in units
of $J_d$. The equilibrium phase diagram of this system has been
extensibly studied~\cite{MaWhRoDe1995,CaStTa2004,RaReTa2006,PiCa2007},
while several dynamical properties at low temperatures were
studied in \cite{SaAlMe1996,ToTaCa1998,StCa1999,GlTaCaMo2003}.
The threshold for the appearance of the stripe phase in this model
is $\delta_c=0.425$~\cite{MaWhRoDe1995,PiCa2007}. For $\delta> \delta_c$
the system presents a sequence of striped ground states,
characterized by a constant width $h$, whose value increases
exponentially with $\delta$ \cite{MaWhRoDe1995,GiLeLi2006}.

In \cite{CaStTa2004} it was shown that in
the range $1 \leq \delta \leq 3$ the system presents a first order
phase transition between a low temperature stripe phase and
a high temperature tetragonal phase with broken translational
and rotational symmetry. In a subsequent work \cite{CaMiStTa2006}
it was shown that for a narrow window around $\delta=2$ the
model shows an intermediate nematic phase, between the stripe and
tetragonal ones, where the system has short range positional
order but long range orientational order.
The nematic phase exists between two critical
    temperatures $T_1<T<T_2$. For finite sizes these are
    actually pseudo critical temperatures $T_1(L)$ and $T_2(L)$,
with finite energy barriers between phases.
The extrapolation to the thermodynamic limit gives $T_1(\infty)
\approx 0.772$ and $T_2(\infty) \approx  0.797$~\cite{CaMiStTa2006}.
For $L \sim 50$
    ($L=48-56$) we can estimate $T_1(L)$ and $T_2(L)$ as the
    average between the positions of the minimum of the cumulant
    and the maximum of the specific heat, so $T_1(50) \approx
    0.78$ and $T_2(50) \approx  0.81$.
    For the range of system sizes considered ($L \sim 50$)
    the energy barriers between the nematic and the tetragonal
    phases are rather small, while the barriers between the stripe
    and the nematic phases are very large~\cite{CaMiStTa2006}.
Hence, under a cooling
    from high temperature no metastable tetragonal states are
    observed, but we see a strong metastability in the nematic
    phase. This can be observed in Fig.19 of
    Ref.\cite{CaMiStTa2006}. Although metastability strongly
resembles a first order phase transition, we were not able to
determine the nature of the stripe-nematic transition unambiguously:
while the nematic order parameter presents a jump at $T_1$ and
the free energy barriers seem to diverge in the thermodynamic limit,
the internal energy is continuous at the transition and the
peak in the specific heat tends to saturate~\cite{CaMiStTa2006}. Further MC simulations showed that the nematic phase is present also for other values of $\delta \neq 2$ in different parts of the phase diagram, mainly around the phase borders between striped states of different widths~\cite{PiCa2007}.

Here we study the out of equilibrium dynamical properties  of the two
dimensional Ising
model with exchange and dipolar interactions described by the
Hamiltonian (\ref{Hamilton1}) with $\delta=2$. We performed quenches of the system
from initial conditions at high temperatures directly to
the stripe phase ($T_f \leq T_1$). Our main result is the identification
of two kinds of processes:
a slow coarsening towards a metastable nematic phase followed by
nucleation of the stripe phase in a background
of the metastable nematic and also
direct coarsening of the stripe phase without an intermediate nucleation
process (see Fig.\ref{quench}). In each realization
the system chooses one of both paths to the final stripe phase with
probabilities that depend on the final temperature of the quench.
We computed these probabilities from Monte Carlo simulations and then
using arguments from homogeneous nucleation theory, finite size scaling
analysis and known growth laws for critical dynamics we were able to
 characterize completely both kinds of processes in the whole
temperature range below the stripe-nematic transition.

We also made slow cooling experiments, similar to those already
discussed in \cite{CaMiStTa2006} and which show strong hysteretic
behavior in the energy. We confirm that hysteresis is associated with a strong
metastability of the nematic phase, while the tetragonal phase
shows negligible metastability in the energy under cooling. Finally,
(quasi-)equilibrium autocorrelation functions obtained for a whole
range of temperatures between the tetragonal and the stripes phases
show slow relaxation characterized by stretched exponential decay.
In the stripe phase the relaxation times obtained from the autocorrelations
exhibit a strong growth which can be well fitted by a
Vogel-Fulcher-Tamman
relation, similar to the behavior of fragile glass formers,
with a relatively high ( $T_0 \simeq 0.45$) divergence temperature.

In Section II we describe the quench experiments and the associated
phenomenology. In Section III the slow cooling experiments and the
behavior of the equilibrium correlations are discussed. In Section
IV we present a discussion of results and brief conclusions.

\section{Relaxation after a quench from high temperature: interplay
between coarsening and nucleation}

We performed quenches from infinite temperature (completely random
initial configuration) down to temperatures $T < T_1$. We also
compared with the case where the initial configuration is taken
from an equilibrated state at temperatures $T \sim 1$. The results
were the same.

\begin{figure*}
\begin{center}
\includegraphics[scale=0.6,angle=-90]{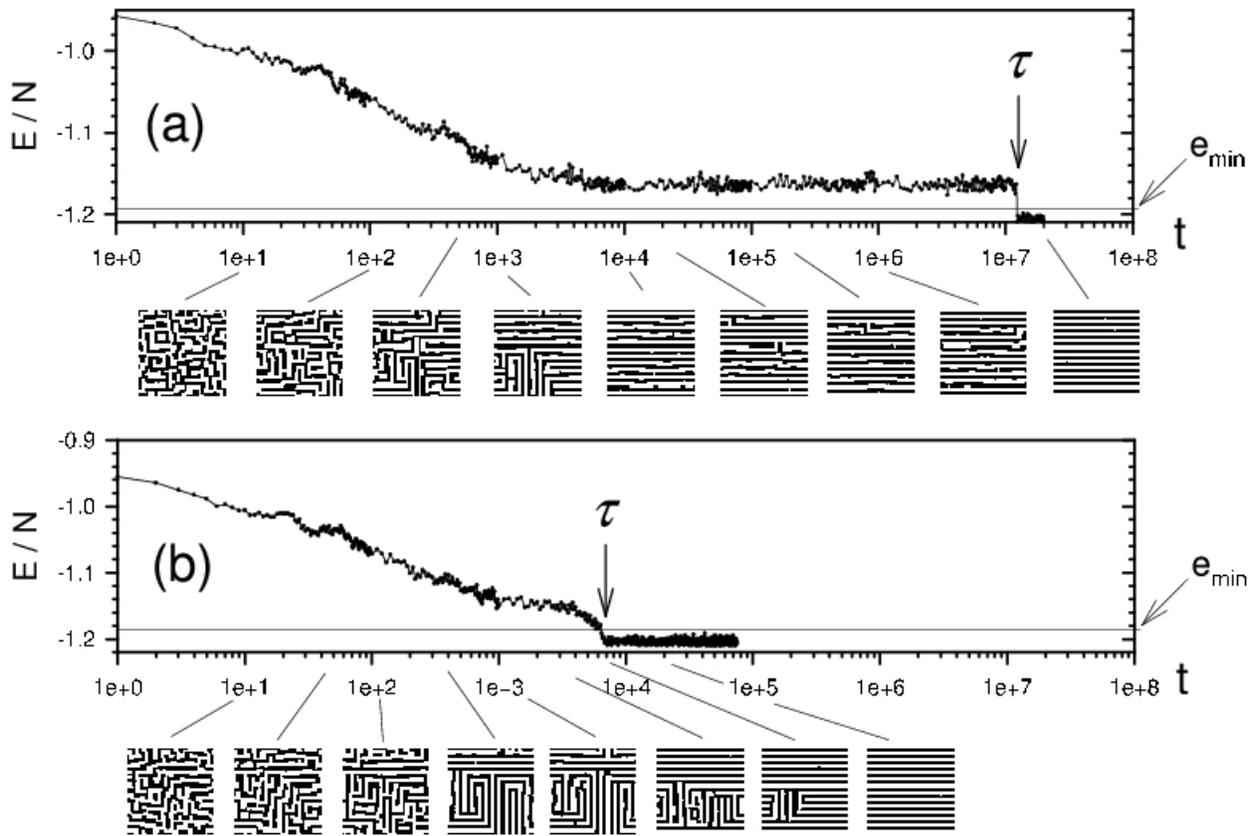}
\caption{\label{quench} Time evolution of the energy per spin in a
single MC run ($\delta=2$, L=52 and $T=0.65$) for two different
realizations of the stochastic noise. Some snapshots of the spin
configurations at different times are shown below the figures. (a)
Competition between nematic domains (coarsening), that leads to a
metastable nematic state, followed by an activated relaxation to
the striped state (equilibration through nucleation); (b)
Competition between nematic and striped domains (coarsening) that
leads directly to equilibration.}
\end{center}
\end{figure*}

The typical relaxation behavior can be appreciated in
Fig.\ref{quench}, where we plotted the time evolution of the
instantaneous energy per spin $E(t)/N$ for two different
realizations of the stochastic noise, together with typical spin
configurations along the evolution. We see that the system can
relax through two different types of mechanisms. In
Fig.\ref{quench}a the system relaxes first into a metastable nematic
state, where it stays during a long time $\tau$, after which
suddenly relaxes to the equilibrium striped state. In other words,
there is a two-step relaxation: first a coarsening process, where
domains  of the nematic phase compete (horizontal and vertical
stripe orientations),
followed by a nucleation of the stable phase (we will verify this
quantitatively later). In Fig.\ref{quench}b the system form
domains of both the stable (striped) and metastable (nematic)
phases that compete and relax directly to the equilibrium state
through a coarsening process. Both types of relaxation processes
can happen with certain probability which depends on the
temperature.  To analyze quantitatively these processes we
computed the equilibration time probability distribution.

The equilibration time $\tau$ is defined as the time (in Monte Carlo
Steps, MCS) that
the system takes to reach the stable, crystal--like striped state
after a quench. The criterium to determine $\tau$ was the
following. For each quench temperature $T < T_1$ we calculated the
equilibrium energy per spin histogram  along a large single MC
path starting from the ground state. This histogram presents two
gaussian--shaped peaks, each one centered at the averages
$u_{str}(T)$ and $u_{dis}(T)$ \cite{CaMiStTa2006}. Between the two
peaks the histogram presents a finite minimum (at least for a
finite system), associated with the free energy barrier that
separates both phases. We also measured the standard deviation of
the low energy peak associated with the striped phase
$\sigma_{str}(T)$ and defined $e_{min}(T)=u_{str}+\sigma_{str}$,
verifying that it is well below the central minimum of the histogram;
$e_{min}$ is almost independent of the system size for $L\geq 32$,
so it can be easily calculated using small system sizes. Then for
each quench we calculated the average energy per spin $e(t)\equiv
\left< H \right>/N$. When $e(t)$ fell bellow $e_{min}$ we stopped
the simulation defining $e(\tau)=e_{min}$ (see Fig.\ref{quench}).
Repeating this procedure we obtained the probability distribution
(normalized frequency) $P(\tau)$ of  the
stochastic variable $\tau$ for different values of $T$ and $L$.

\begin{figure}
\begin{center}
\includegraphics[scale=0.34,angle=-90]{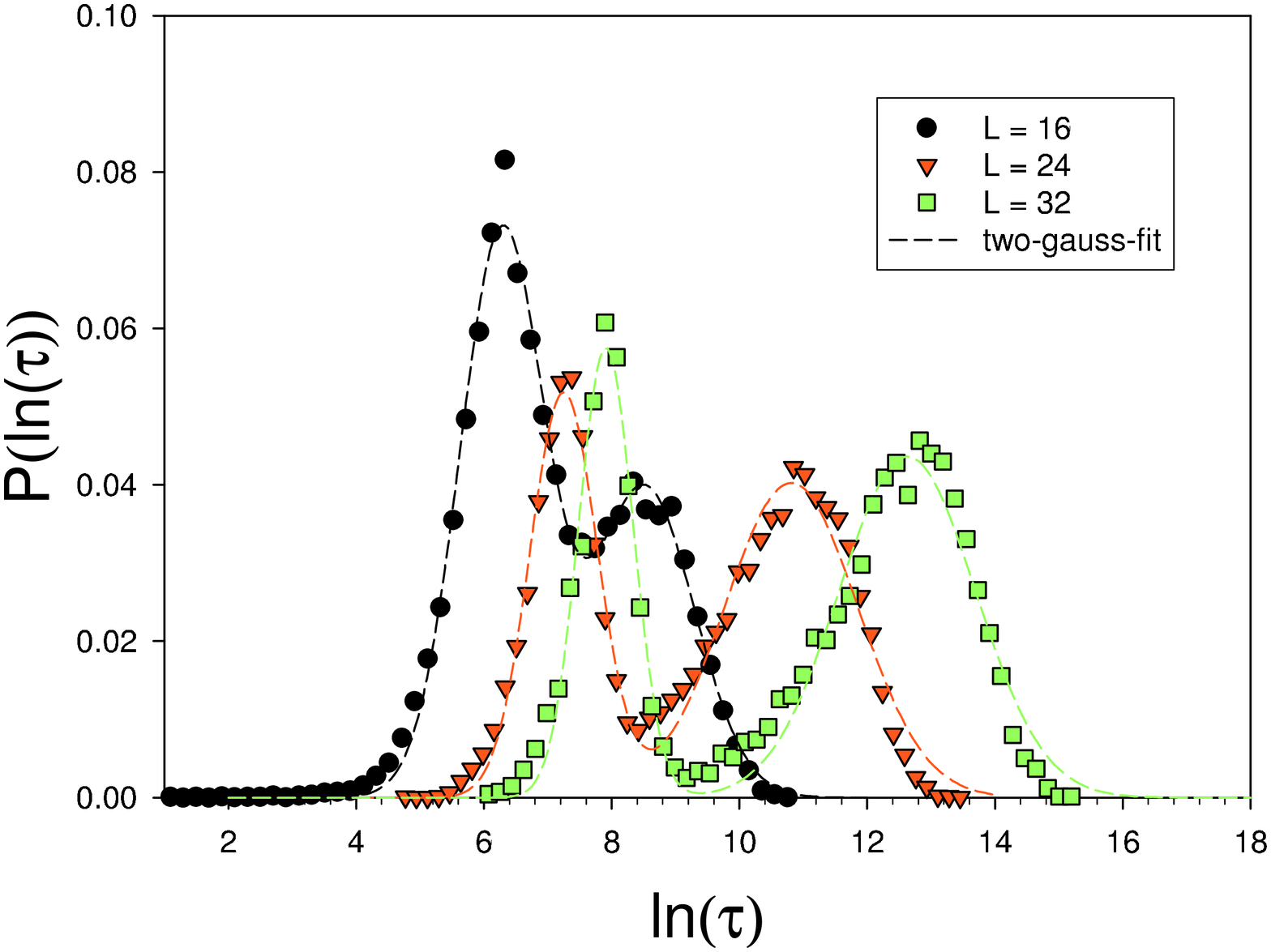}
\caption{\label{histop1} (Color online) Normalized frequency
$P(\tau)$ vs $\ln{\tau}$ for $\delta=2$, $T=0.65$ and different system sizes.
The lines correspond to a fitting with a superposition of two
log-normal distributions. Sample size for the histograms were
$\sim 2 \times 10^4$.}
\end{center}
\end{figure}

\begin{figure}
\begin{center}
\includegraphics[scale=0.36,angle=-90]{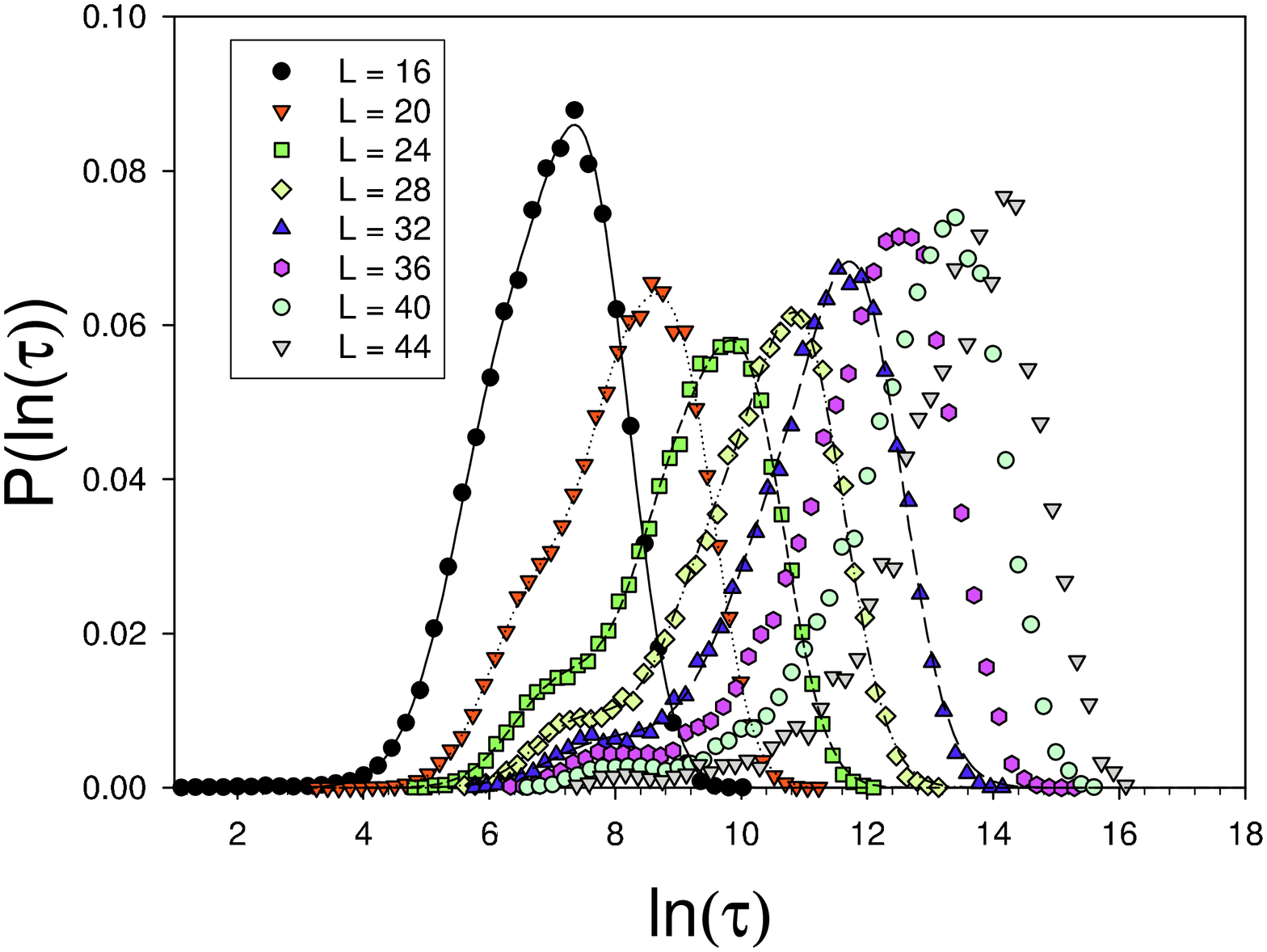}
\caption{\label{histop2} (Color online) Normalized frequency
$P(\tau)$ vs $\ln{\tau}$ for $\delta=2$, $T=0.77$ and different system sizes.
The lines correspond to a fitting with a superposition of two
log-normal distributions Eqs.(\ref{fit}). Sample sizes for the
histograms run between $\sim 4 \times 10^3$ ($L=44$) to $\sim 5
\times 10^4$ ($L=16$).)}
\end{center}
\end{figure}

\begin{figure}
\begin{center}
\includegraphics[scale=0.36,angle=-90]{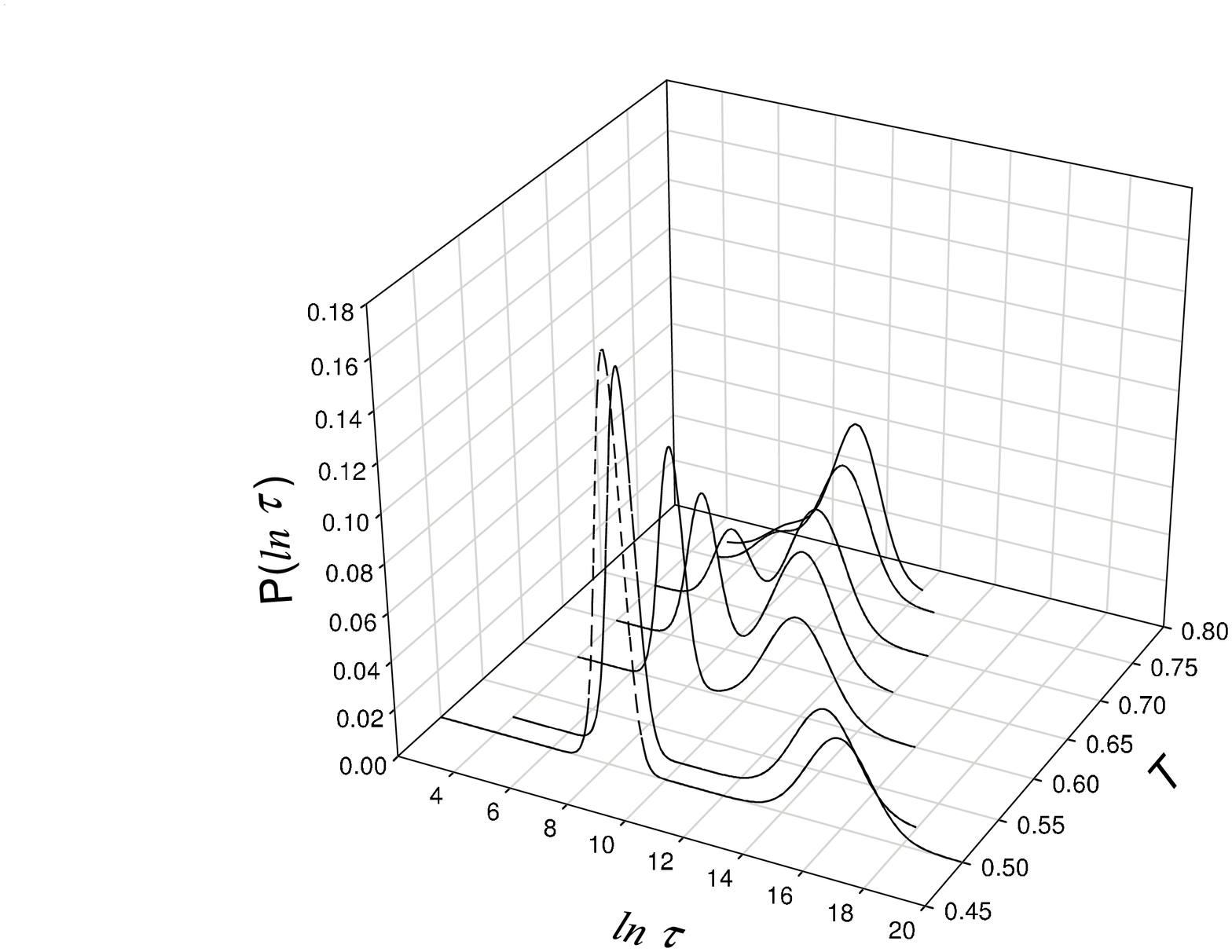}
\caption{\label{histop3}  Fitted probability distribution
$P(\tau)$ as a function of $\ln{\tau}$ and $T$ for $\delta=2$ and $L=24$.}
\end{center}
\end{figure}

The above mentioned relaxation mechanisms are reflected in a
characteristic two-peak structure in $P(\tau)$, each one centered at typical
values corresponding to well different time scales, as shown
in Figs.\ref{histop1} and \ref{histop2}. This
probability distribution can be very well fitted using a
superposition of two log--normal functions (see Figs.\ref{histop1}
and \ref{histop2}) $P(\ln{\tau}) = P_1(\ln{\tau})+P_2(\ln{\tau})$
with
\begin{equation}
P_i(\ln{\tau}) = A_i \;
\exp{\left(\frac{(\ln{\tau}-\mu_i)^2}{2\;\sigma_i^2}\right)}
\;\;\;\ i=1,2 \label{fit}
\end{equation}
\noindent where $A_i$ are normalization constants. Using these
fittings we can see in Fig. \ref{histop3} that the characteristic
time scales of both peaks increases as $T$ decreases.

These fittings also allowed us to estimate the averages $\left<
\tau \right>_i = \exp{\left(\mu_i + \sigma_i^2/2 \right)}$,
$i=1,2$, associated with each process. Both averages increase with
$L$, so we can use finite size scaling analysis to determine the
nature of the associated processes.

\subsection{Finite size scaling of the equilibration times}

\begin{figure}
\vspace{0.5cm}
\begin{center}
\includegraphics[scale=0.36,angle=-90]{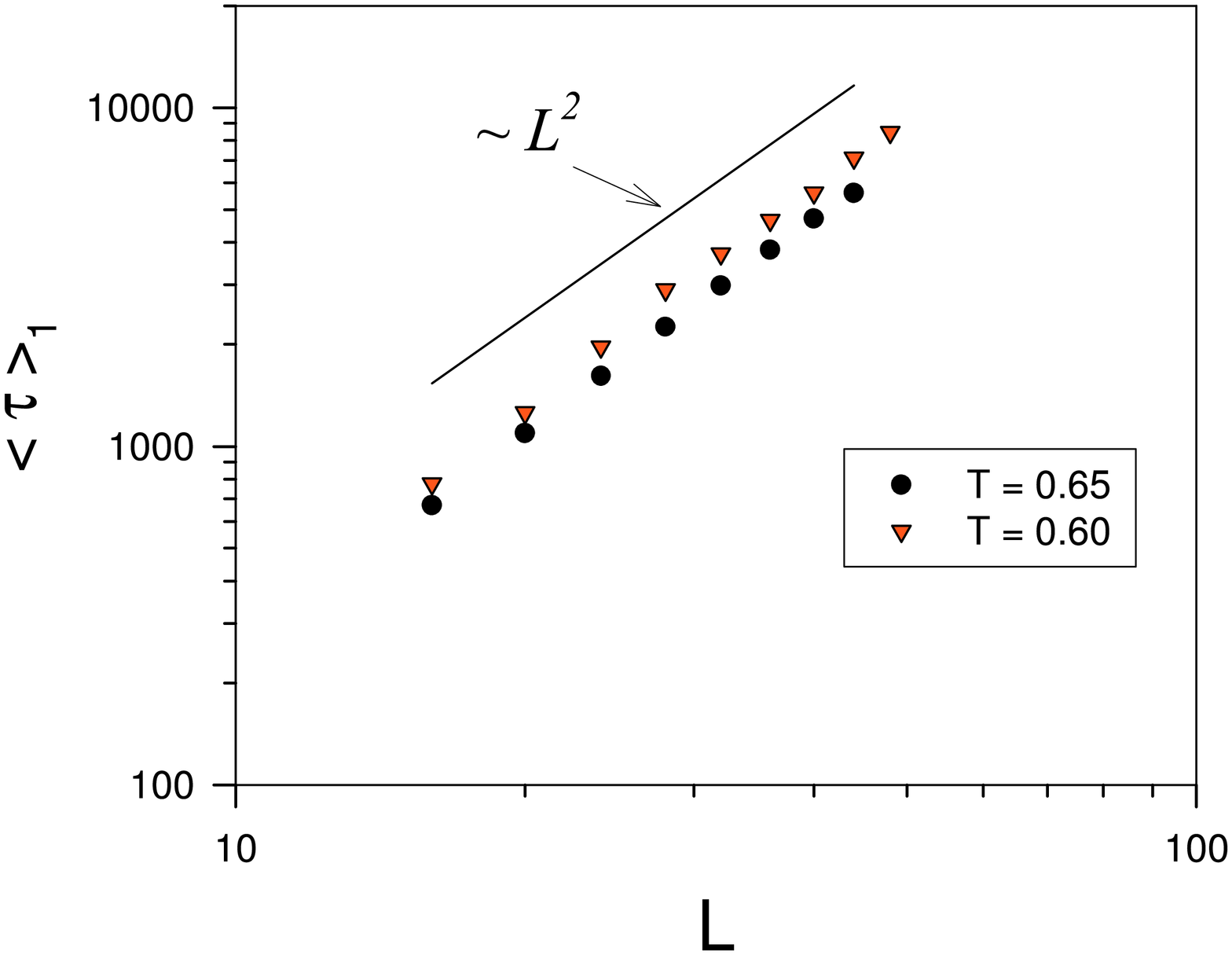}
\caption{\label{tau1}  (Color online)  $\left< \tau \right>_1$ vs.
L from two log-normal fittings Eqs.(\ref{fit})}
\end{center}
\end{figure}

Suppose that the relaxation is completely governed by a
competition between ground state domains, that is, by a curvature
driven relaxation of the domain walls. The  excess of energy
respect to the ground state is given by $\delta e = \gamma n$,
$\gamma$ being the "surface" tension per unit length of the domain
walls and $n$ the fraction of spins belonging to them; $n$ is
approximately given by $n \simeq b N_d(t) l(t)/N$, where $N_d(t)$
is the average
number of domains, $l(t)$ is the average linear size of the
domains and $b$ is a geometrical factor. Using  $N_d \propto
N/l^2(t)$ and grouping all these factors we get $\delta e \sim
l^{-1}(t)$. For $\delta=2$ the energy distribution of the striped
phase is very narrow and lies very near the ground state, so the
criterium $e(\tau)=e_{min}$ implies $\delta e \ll 1$, condition
that is attained when $l(\tau)\sim L$. Since for a heat bath
dynamics with non-conserved order parameter (class A system)
we can assume $l(t) \sim t^{1/2}$ we
expect the scaling $\left< \tau \right> \sim L^2$. We verified
that $\left< \tau \right>_1 \sim L^2$ for a wide range of values
of $T$ and $L$ (see an example in  Fig.\ref{tau1}), thus
confirming that $P_1(\tau)$ is associated with a simple coarsening
process.

\begin{figure}
\begin{center}
\includegraphics[scale=0.36,angle=-90]{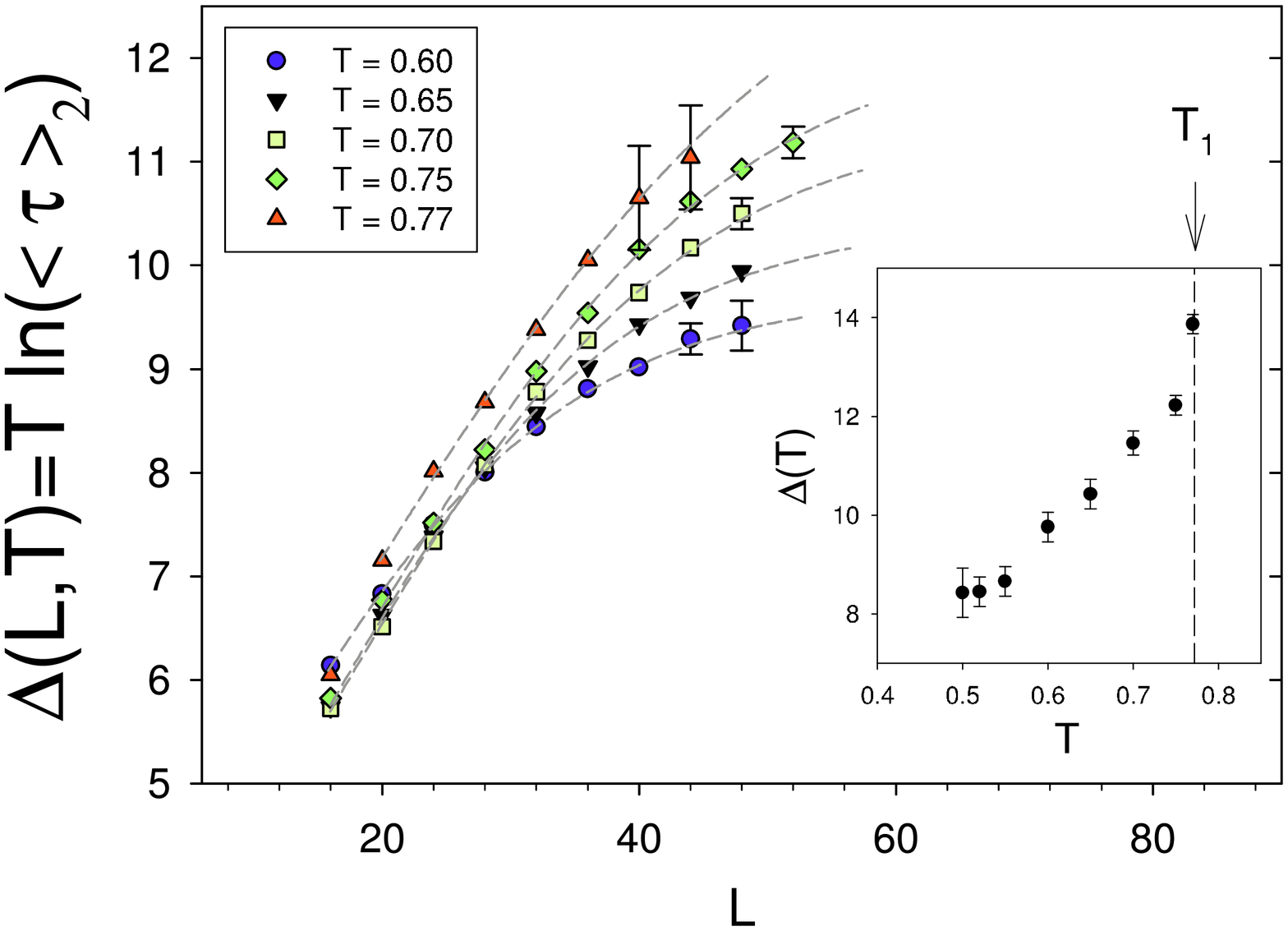}
\caption{\label{tau2}  (Color online)  Estimated free energy
barrier $\Delta(L,T)\approx T\, \ln{\left(\left< \tau \right>_2
\right)}$ vs. $L$ for different temperatures. The dashed lines
correspond to a sigmoidal fitting. The inset shows the
extrapolated free energy barrier $\Delta(T)=\Delta(\infty,T)$ vs.
$T$. The error bars are shown only when larger than the symbol
size.}
\end{center}
\end{figure}

Now suppose that relaxation is governed by an homogeneous
nucleation process. According to classical nucleation theory
$\left< \tau \right>$ follows approximately an Arrhenius law $
\left< \tau \right> = \tau_0\; \exp{(\Delta(L,T)/T)}$,
$\Delta(L,T)$ being the height of the free energy barrier to
nucleation. In a finite two--dimensional system
\cite{LeKo1991} $\Delta(T_m) \sim L$, $T_m$ being the melting
temperature; in our case $T_m=T_1$. In an infinite system below
the melting point, classical nucleation theory  predicts that
$\Delta(T)\equiv \Delta(\infty,T)$ is given by the excess of free
energy of a critical droplet with average linear size $\xi(T)$ of
the ordered (striped) phase inside the disordered (nematic) metastable state.  Thus,
if $L \gg \xi(T)$ the free energy barrier is expected to be almost
independent of the system size when $T<T_m$. But, if  $L <
\xi(T)$, the droplet never reaches the critical size and the
excess of free energy is dominated by surface tension. Hence,  the
transition to the crystal phase will occur when the crystal
droplet reaches the system size and therefore  we expect
$\Delta(L,T) \sim L$ and $\ln{\left< \tau \right>} \sim L/T$.
Fig.\ref{tau2} shows that $\left< \tau \right>_2$ follows such
scaling, suggesting its identification with the average nucleation
time $\tau_{nucl}$. However, due to the procedure we used to
obtain it, $\left< \tau \right>_2$ actually corresponds to the
crystallization time, which is not necessarily equal to
$\tau_{nucl}$, because after the creation of a critical droplet it
follows a growth process until the whole system reaches the
crystal phase (for the rather small system sizes we considered we
can assume that there is only one critical droplet). Hence,
$\left< \tau \right>_2 = \tau_{nucl}+\tau_{growth}$. However, we
can expect $\tau_{growth}\sim L^n$, so $\tau_{nucl} \gg
\tau_{growth}$, at least as far as $L$ is large enough that we are
in the scaling regime, but small enough so that the free energy
barrier still depends on $L$. Under these conditions $\left< \tau
\right>_2$ give us a good approximation of $\tau_{nucl}$ and we
can estimate $\Delta(L,T) \approx T\, \ln{\left< \tau \right>_2}$
up to logarithmic corrections. Then, by fitting the curves of
Fig.\ref{tau2} using a sigmoidal function $ \Delta(L,T) =
\Delta(T)/\left(1+\exp{(L^*(T)-L)/c(T)}\right)$ ($L^*(T)$,
$\Delta(T)$  and $c(T)$ are fitting parameters) we can extrapolate
$\Delta(T)=\lim_{L\rightarrow\infty} \Delta(L,T)$. The
extrapolated curve $\Delta(T)$ is shown in the inset of
Fig.\ref{tau2}. We see that $\Delta(T)$ grows strongly as
$T\rightarrow T_1^-$, while it appears to present a minimum around
$T=0.5$, or at least to saturate in a constant value as
$T\rightarrow 0$. This is compatible with the results in
\cite{CaMiStTa2006} obtained from a different analysis where
it was found a divergent free energy barrier
at the transition temperature $T_1$ in the thermodynamic limit.
For quench temperatures $T<0.5$ very large
relaxation times make it very difficult to obtain reliable
numerical results. For future analysis we list in Table
\ref{tabla1} the approximated fitted values of $\left< \tau
\right>_2$ for different values of $L$ and $T$.

\begin{figure}
\begin{center}
\includegraphics[scale=0.36,angle=-90]{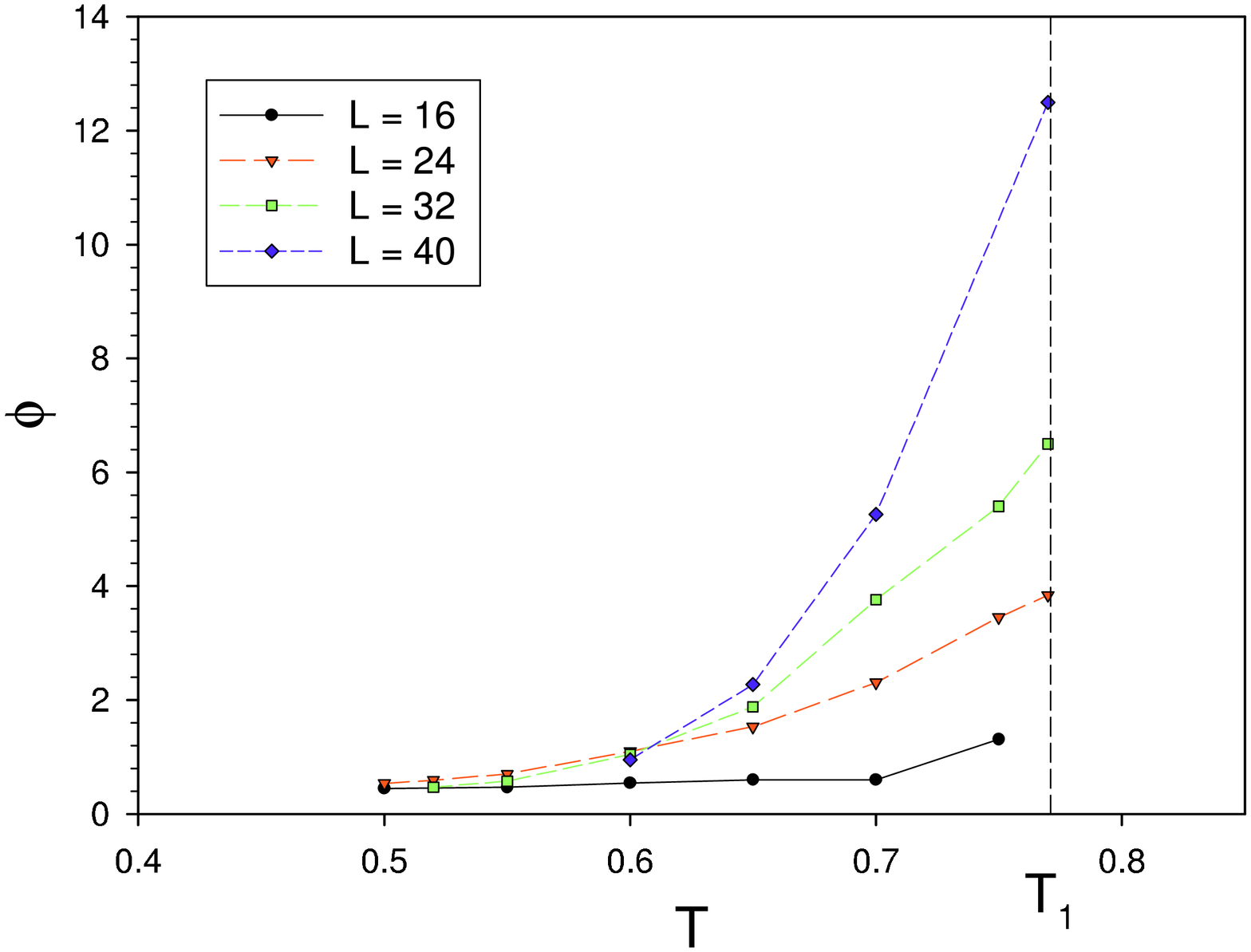}
\caption{\label{fi} (Color online) Relative probability
Eq.(\ref{phi}) vs. $T$ for $\delta=2$ and different values of $L$.}
\end{center}
\end{figure}

\begin{table*}
\begin{center}
\begin{tabular}{|c|c|c|c|c|c|}
\hline
\mbox{ } & $T=0.60$ & $T=0.65$ & $T=0.70$ & $T=0.75$ & $T=0.77$\\
\hline
 $L=32$ & $\sim 1.3 \times 10^6$ & $\sim 5.4 \times 10^5$ & $\sim 2.8 \times 10^5$ & $\sim 1.6 \times 10^5$ & $\sim 2.0 \times 10^5$\\
\hline
 $L=36$ & $\sim 2.4 \times 10^6$ & $\sim 1.1 \times 10^6$ & $\sim 5.7 \times 10^5$ & $\sim 3.3 \times 10^5$ & $\sim 4.7 \times 10^5$\\
\hline
 $L=40$ & $\sim 3.4 \times 10^6$ & $\sim 2.0 \times 10^6$ & $\sim 1.1 \times 10^6$ & $\sim 8.0 \times 10^5$ & $\sim 1.0 \times 10^6$\\
\hline
 $L=44$ & $\sim 5.3 \times 10^6$ & $\sim 3.0 \times 10^6$ & $\sim 2.0 \times 10^6$ & $\sim 1.4 \times 10^6$ & $\sim 1.6 \times 10^6$\\
\hline
 $L=48$ & $\sim 6.6 \times 10^6$ & $\sim 4.4 \times 10^6$ & $\sim 3.3 \times 10^6$ & $\sim 2.1 \times 10^6$ & \mbox{ }\\
\hline
\end{tabular}
\end{center}
\caption{$\left< \tau \right>_2$ for different values of $L$ and
$T$.} \label{tabla1}
\end{table*}

We also calculated the relative probability of occurrence of
nucleation and coarsening processes

\begin{equation}
\phi(T) \equiv \frac{P_{nucl}}{P_{coars}} =
\frac{\int_{-\infty}^\infty P_2(\ln{\tau})
d(\ln{\tau})}{\int_{-\infty}^\infty P_1(\ln{\tau}) d(\ln{\tau})} =
\frac{\sigma_2\; A_2}{\sigma_1\; A_1} \label{phi}
\end{equation}

\noindent in the range $0.5 \leq T<T_1$. In Fig.\ref{fi} we see
that $\phi(T)$ decreases monotonously with $T$. It seems to diverge
as $T\rightarrow T_1^-$ and $L\rightarrow\infty$, while becoming
almost independent of $L$ for $T<0.6$, saturating in a
constant value $\phi \approx 0.5$. This shows that nucleation
dominates the relaxation near the melting point, while coarsening
becomes dominant for deep quenches, although there is always a
finite and relatively high ($\sim 1/3$) probability of
nucleation.

\section{Relaxation after a slow cooling}

\subsection{Energy}

\begin{figure}
\begin{center}
\includegraphics[scale=0.34,angle=-90]{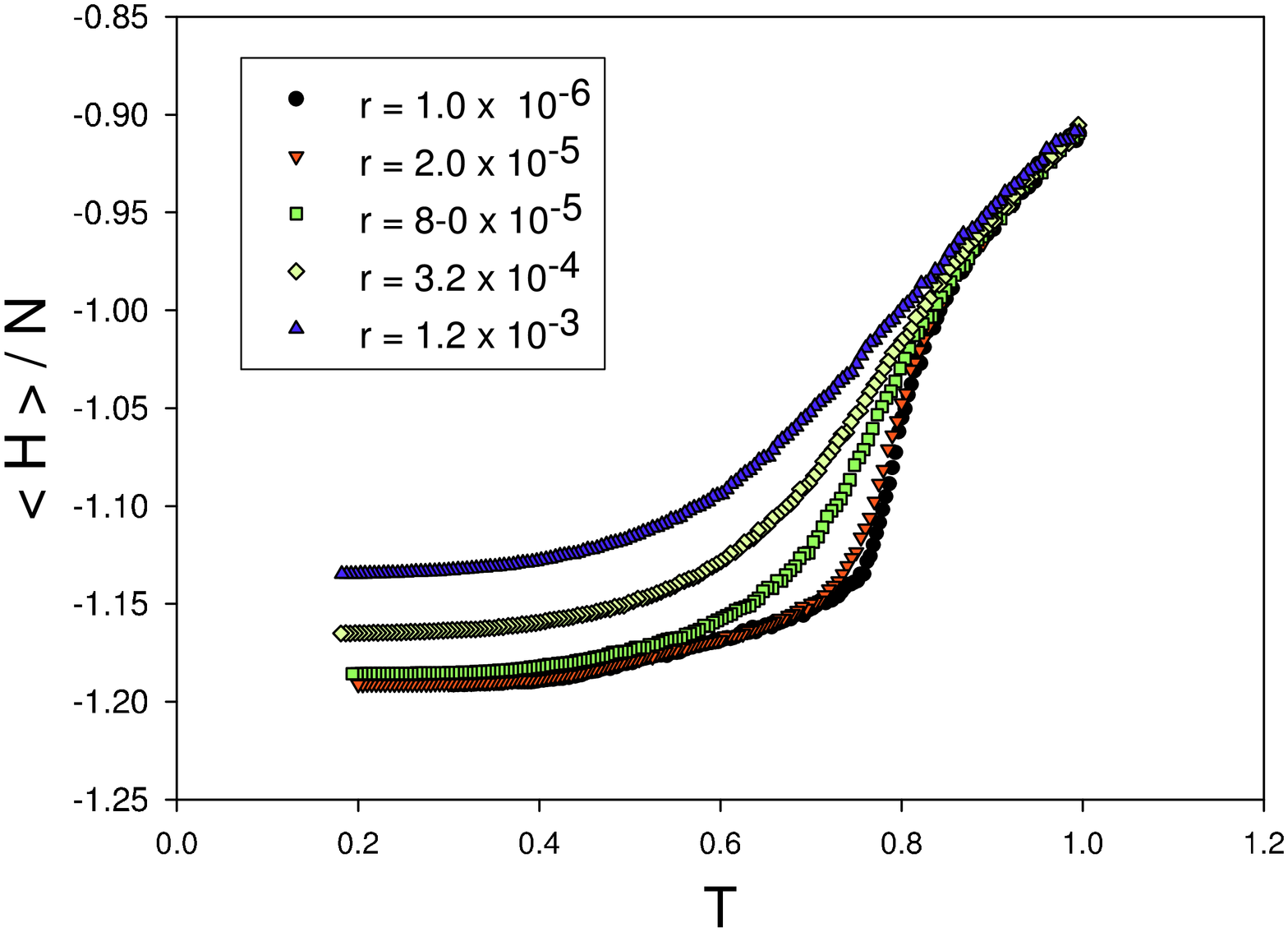}
\caption{\label{energia1}  (Color online)  Average energy per spin in a  cooling
from $T=0.9$ ($\delta=2 and $$L=48$) for different cooling rates.}
\end{center}
\end{figure}

\begin{figure}
\begin{center}
\includegraphics[scale=0.34,angle=-90]{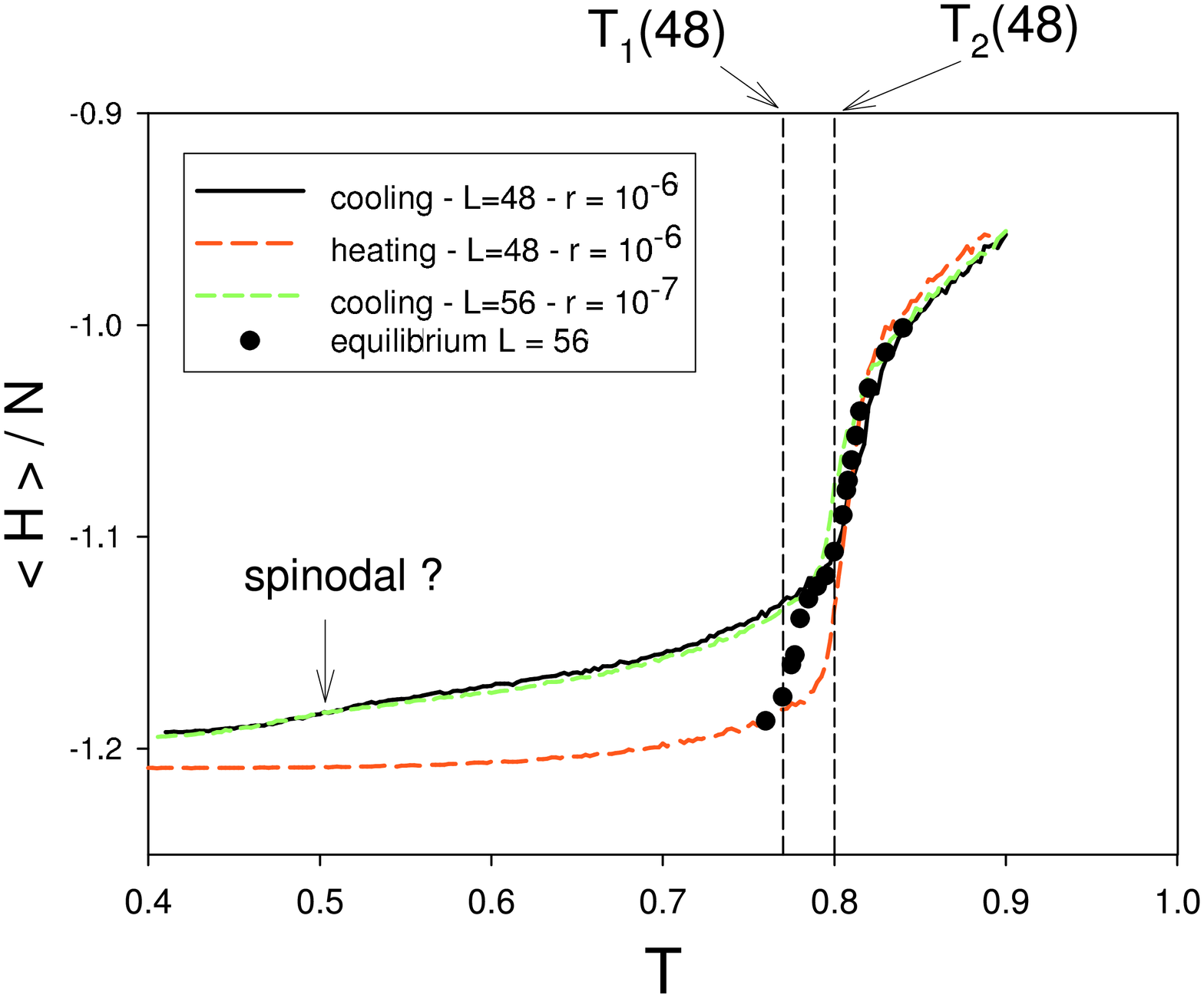}
\caption{\label{energia2}   (Color online) Average energy per spin cooling-heating
curves ($\delta=2$ and  $L=48$).}
\end{center}
\end{figure}

\begin{figure}
\begin{center}
\includegraphics[scale=0.34]{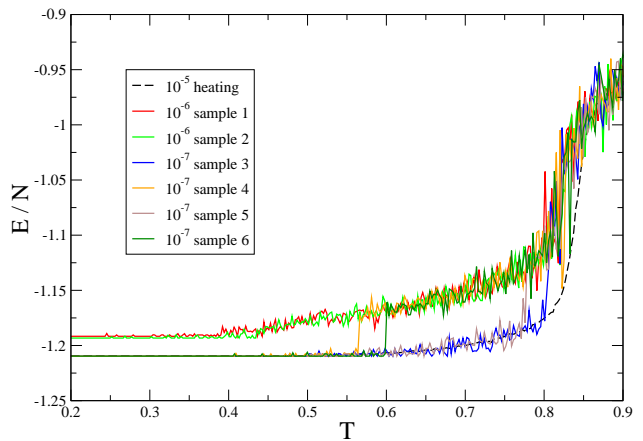}
\caption{\label{energia3}  (Color online) Energy per spin  cooling curves:
individual realizations ($\delta=2$ and  $L=48$).}
\end{center}
\end{figure}

We first analyze the behavior of the average energy per spin
during a slow cooling; that is, we first equilibrate the system at
some temperature $T_0 > T_2$ and then decrease the temperature
down to a final value $T_f  \ll T_1$ according to the linear
protocol $T(t) = T_0 - r\, t$, where $r$ is the cooling rate, $t$
is in MCS and the average is taken over several runs between $T_0$
and $T_f$. In Fig.\ref{energia1} we show an example. We see that
below some value of $r$ the energy curve becomes almost
independent of $r$. In  Fig.\ref{energia2} we compare the cooling
curves with the corresponding heating curve; this last curve is
calculated by performing a linear heating from the ground state,
starting at a very low temperature. We also compare the curves
with the equilibrium values obtained in Ref.\cite{CaMiStTa2006}.
These results verify that (for this range of  system sizes)
metastability in the tetragonal-nematic transition  is negligible,
while there is a strong metastability associated with the
stripe-nematic transition. Notice that the metastable cooling
curve has an inflection point around $T=0.5$ and falls down;
although it does not reach the equilibrium value for $T <0.5$,
this inflection point suggests the existence of a spinodal
temperature around $T = 0.5$. Moreover, from Table \ref{tabla1} we
see that for $L \sim 50$ the nucleation time is $\tau_{nucl} \sim
4 \times 10^6$ MCS; then, for a cooling rate $r=10^{-7}$, the time
elapsed from the instant at which the temperature $T=T_1$ to the
instant corresponding to a final temperature around $T=0.5$ will
be $t = \Delta T/r \sim \tau_{nucl}$. Therefore, for that range of
temperatures we would expect that in a large fraction of the
realizations the system would have already equilibrated. Indeed,
the results of Fig.\ref{energia3} verify this assumption.
Moreover, the tracking of typical spin configurations for those
realizations were the system did not equilibrate below $T=0.5$
shows that actually the system is no longer in a nematic state;
instead of that, it got stuck in another metastable state, with a
mixture of stripes  with widths $h=2$ and $h=3$. Actually the
$h=3$ striped phase becomes metastable precisely for that range of
temperatures\cite{PiCa2007}. This behavior reinforces the
hypothesis of the existence of a spinodal temperature $T_{sp}
\approx 0.5$. Also the behavior of the orientational order
parameter observed in Fig.19 of Ref.\cite{CaMiStTa2006} supports
it.

\subsection{Correlations}

We now analyze the two-times correlation function

\begin{equation}
C(t_w,t_w+t) =\frac{1}{N} \sum_i \left< S_i(t_w)\; S_i(t_w+t)
\right> \label{correlation}
\end{equation}

\noindent after a cooling from high temperature to a fixed
temperature $T< T_2$. In this case we performed the cooling with a
ladder protocol, that is, we reduced the temperature by steps of
$\Delta T = 0.01$, letting the system to equilibrate during $10^4$
MCS at each intermediate temperature. On the average, this
corresponds to a cooling rate $r=10^{-6}$. Once we arrived to the
measuring temperature (and after another $10^4$ MCS) we set $t=0$
and we saved the initial configuration. The autocorrelations
(\ref{correlation}) were then measured as a function of $t$ and $t_w$
and the whole curve was averaged over different realizations of
the stochastic noise, starting always from the same initial
configuration (this is to save CPU time). Some check repeating the
calculation starting from different initial configurations (all
obtained with the same cooling protocol) showed no difference with
the previous one.

\begin{figure}
\begin{center}
\includegraphics[scale=0.32,angle=-90]{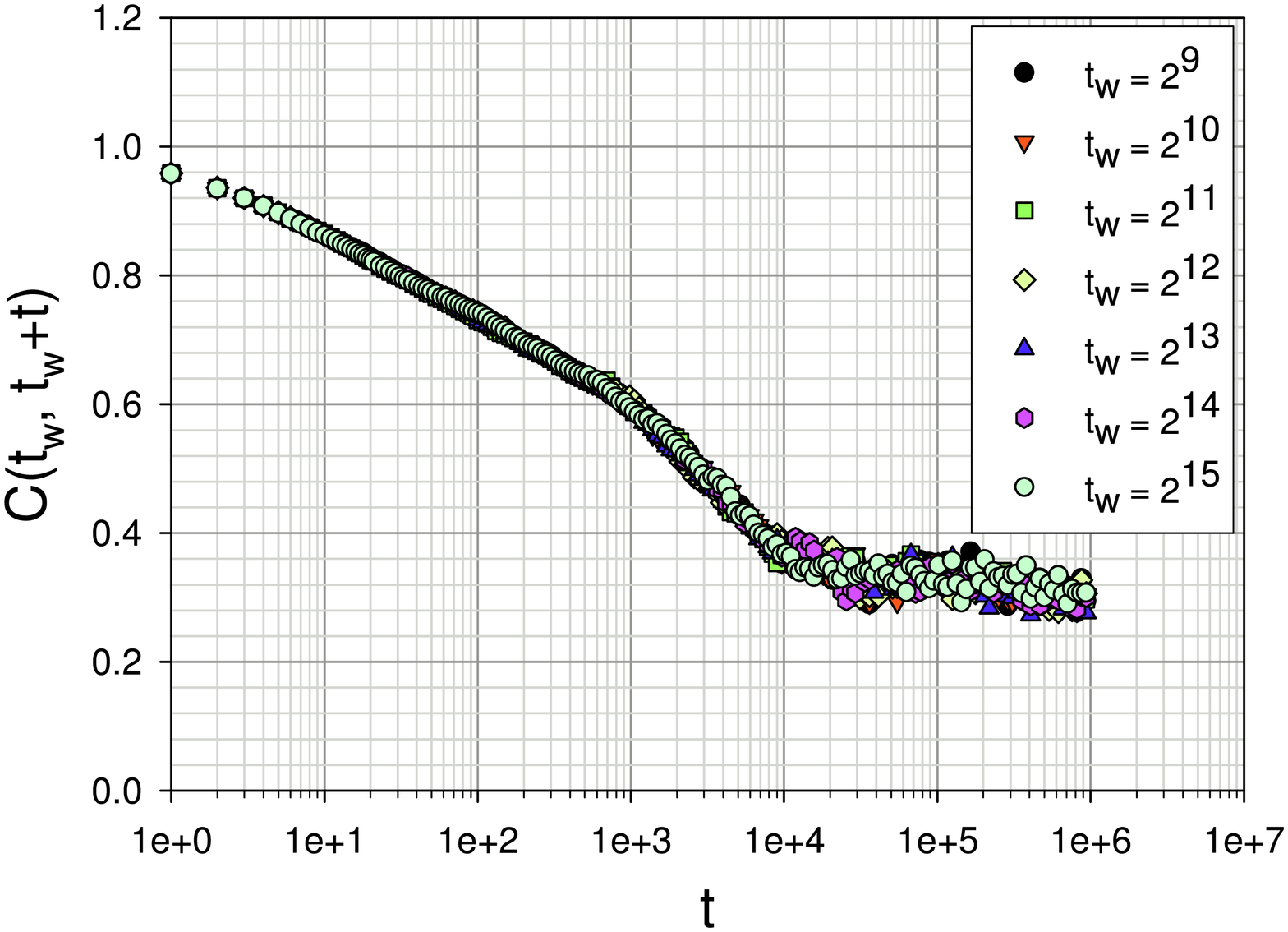}
\caption{\label{corre1}  (Color online)  Two-times correlation function
$C(t_w,t_w+t)$ as a function of $t$ for $\delta=2$, $L=48$ and
different waiting times $t_w$,  after a quasi-static cooling at
$T=0.78$. Sample size for average was $150$}
\end{center}
\end{figure}

First of all we analyzed a cooling down to a temperature $T_1(L) <
T < T_2(L)$, which corresponds to the stable region of the nematic
phase. From  Fig.\ref{corre1} we see that $C(t_w,t_w+t) = C(t)$,
as expected in a stable phase. We also noticed that the
correlations do not decay to zero but to
a plateau $C_{plateau} \approx 0.37$. We will
discuss the meaning of that plateau later.

\begin{figure}
\begin{center}
\includegraphics[scale=0.32,angle=-90]{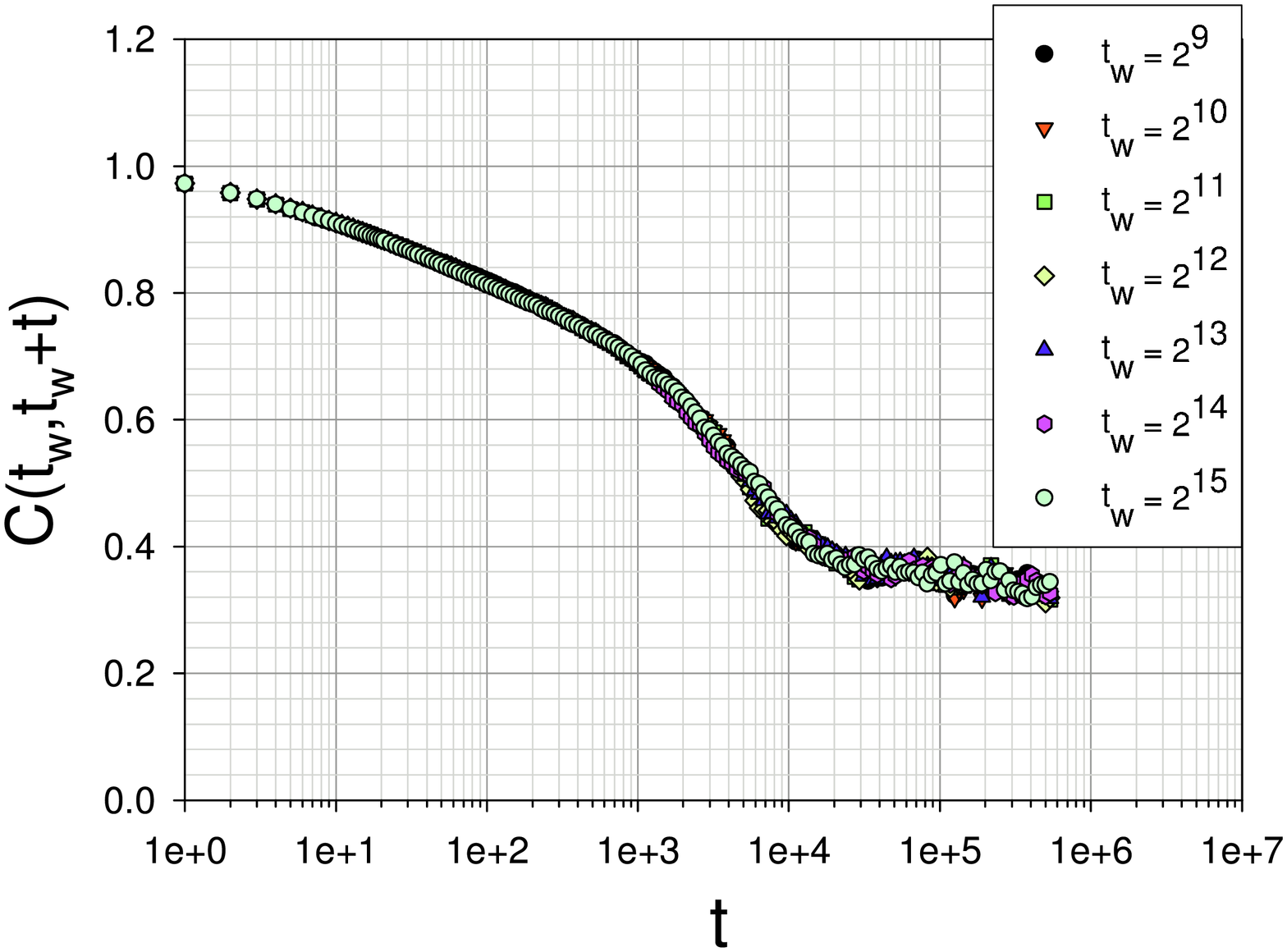}
\caption{\label{corre2}  (Color online) Two-times correlation function
$C(t_w,t_w+t)$ as a function of $t$ for $\delta=2$, $L=48$ and
different waiting times $t_w$,  after a quasi-static cooling at
$T=0.7$. Sample size for average was $350$}
\end{center}
\end{figure}

In Fig.\ref{corre2} we show the same calculation, but for a
temperature $T<T_1(L)$. Both $t$ and $t_w$ were chosen such that
$t+t_w \ll \tau_{nucl}$. We see that again $C(t_w,t_w+t) = C(t)$,
consistently with the expected quasi-stationary nature of a
metastable state. We also see that the correlation also displays
the characteristic plateau observed in the stable nematic state.

\begin{figure}
\begin{center}
\includegraphics[scale=0.32,angle=-90]{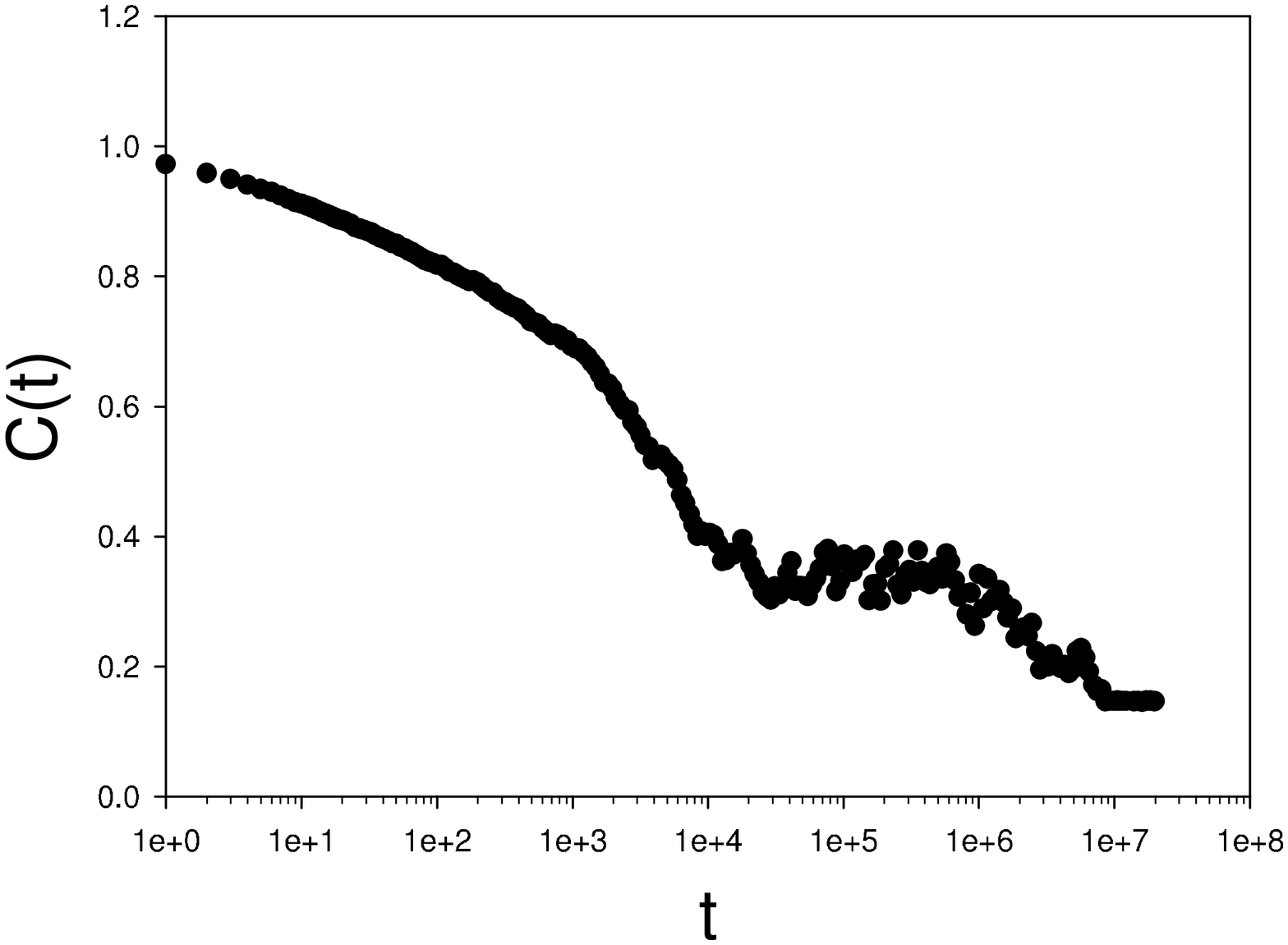}
\caption{\label{corre3} Correlation function
$C(t)$ as a function of $t$ for $\delta=2$, $L=48$ and
different waiting times $t_w$,  after a quasi-static cooling at
$T=0.7$. Sample size for average was $80$}
\end{center}
\end{figure}

In Fig.\ref{corre3} we show $C(t)$ for $T=0.7$ and $L=48$ in a very
long run with $t$ up to $2 \times 10^7$ MCS. Comparing with table \ref{tabla1}
we see that the second decay in the correlation coincides with the
average nucleation time.

\begin{figure}
\begin{center}
\includegraphics[scale=0.37,angle=-90]{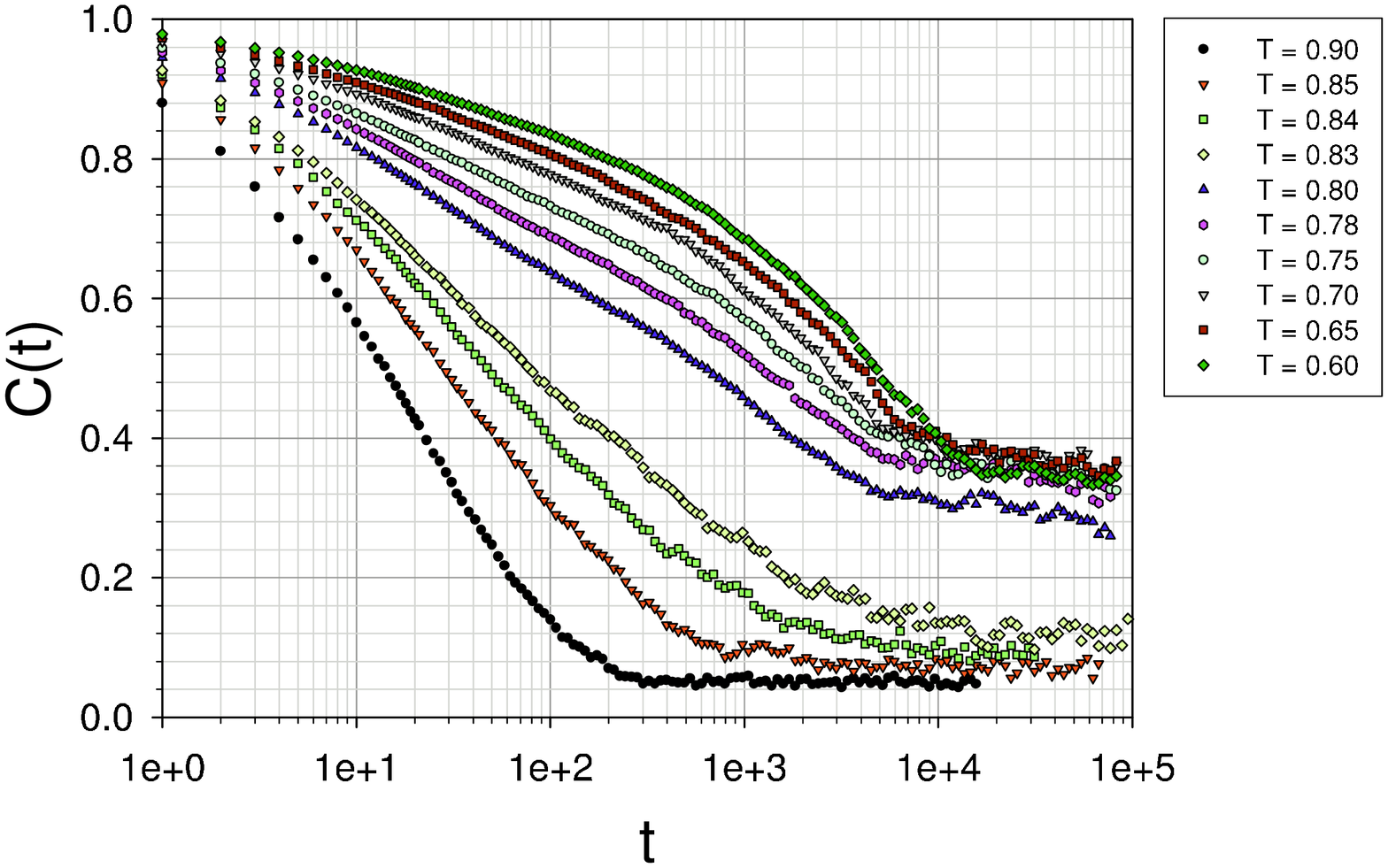}
\caption{\label{corre4}  (Color online)  Correlation function $C(t)$ as a function of $t$ for $\delta=2$ and $L=36$ after a quasi-static cooling at different final temperatures.}
\end{center}
\end{figure}

In Fig.\ref{corre4} we show $C(t)$ for $L=36$ and different cooling
temperatures. We see that the plateau appears for any temperature, but
becomes higher in the nematic case.

\begin{figure}
\begin{center}
\includegraphics[scale=0.32,angle=-90]{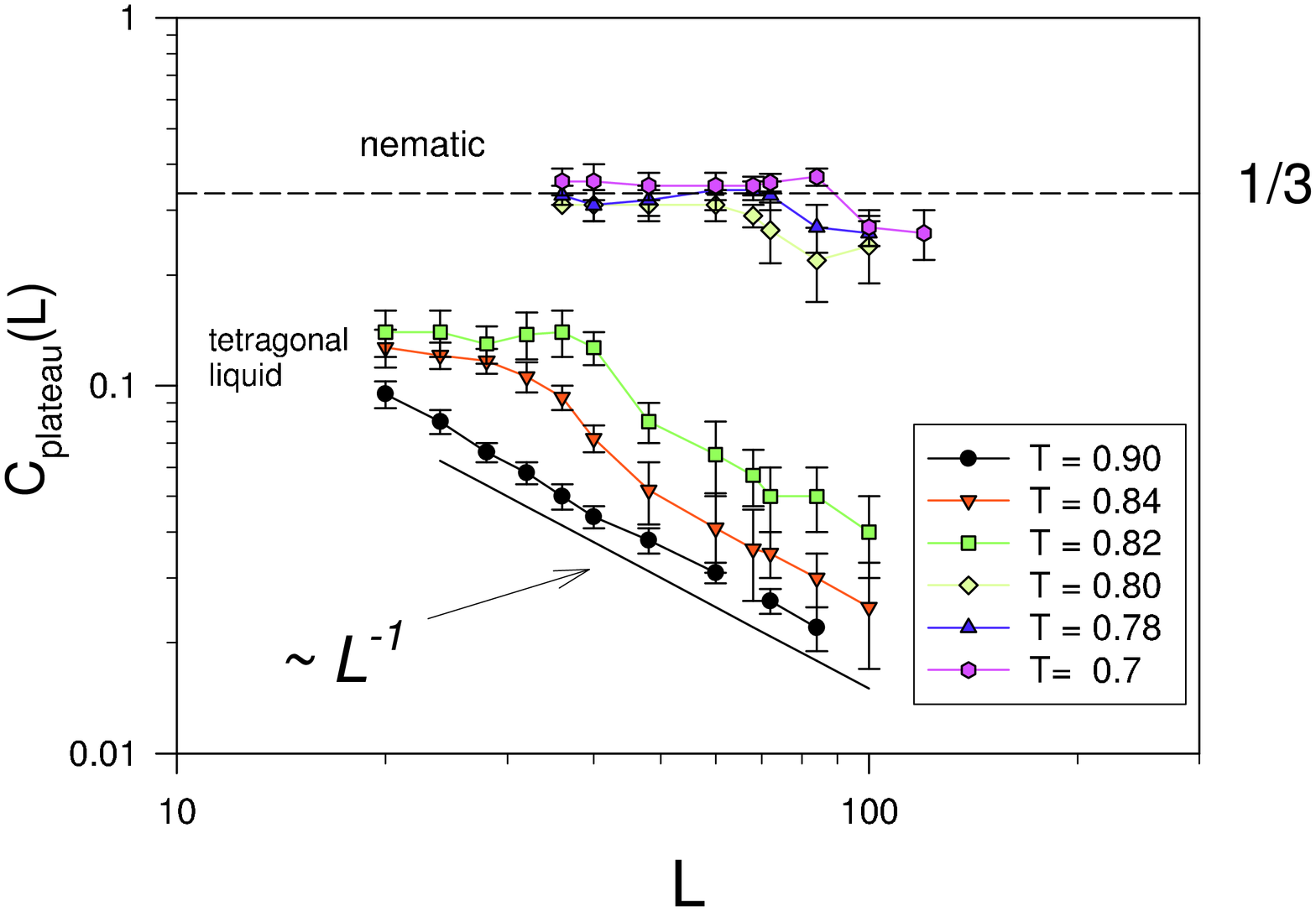}
\caption{\label{plateau}  (Color online) Correlation function plateau
$C_{plateau}$ as a function of $L$ for $\delta=2$ and
different temperatures.}
\end{center}
\end{figure}

What is the origin of the plateau? In Fig.\ref{plateau} we show $C_{plateau}$ as
a function of $L$ for different temperatures. For temperatures in the
tetragonal liquid region, we see that for large enough sizes the plateau decays
as $1/\sqrt{N}$, as expected in a disordered state. However, for small sizes
the plateau is independent of $L$ and the crossover size increases as the
temperature approaches $T_2$. Once we are in the nematic phase (either stable
or metastable) the plateau for small sizes becomes independent of $T$ (at least
for a range of temperatures where the relative probability of nucleation is high,
see Fig. \ref{fi})  and it
is roughly equal to $1/3$. Since in a stationary state the unconnected
correlations we are considering here behave as $C(t) \sim \frac{1}{N} \sum_i
\left<  S_i \right>^2$, this result shows that in the nematic state one third
of the spins remain in a frozen state  for small system sizes.
 Assuming then that  the nematic state in a finite system is characterized by this
residual correlation, we can define a normalized correlation function $C'(t)$ as

\begin{figure}
\begin{center}
\includegraphics[scale=0.35,angle=-90]{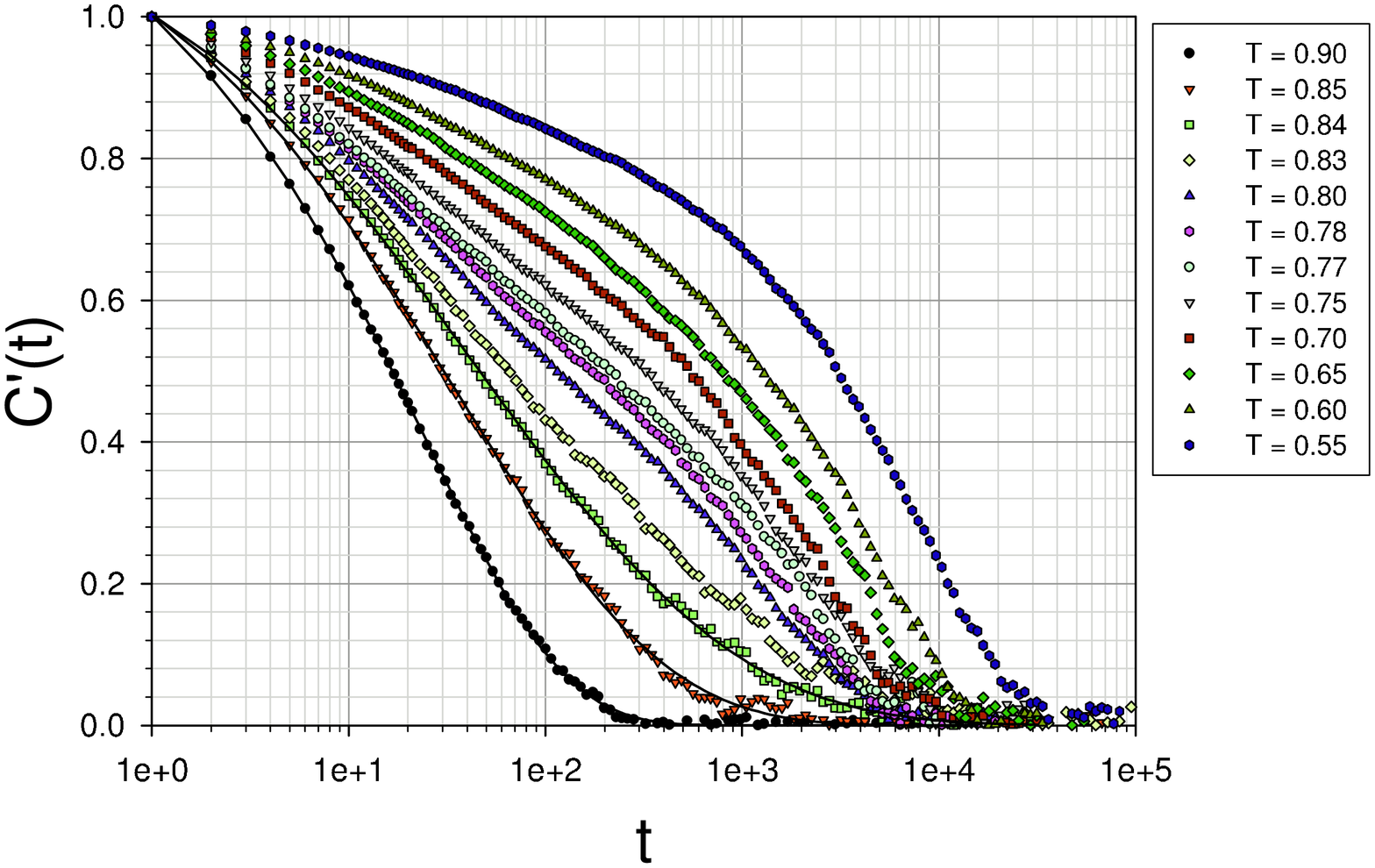}
\caption{\label{corre5}  (Color online) Normalized correlation function Eq.(\ref{Cp}) as a function of $t$ for $\delta=2$ adn $L=36$, after a quasi-static cooling at different temperatures. The continuous lines for temperatures above $T_2$ correspond to  fittings using Eq.(\ref{fit1}).}
\end{center}
\end{figure}

\begin{equation}
C'(t) \equiv \frac{\left(C(t)-C_{plateau} \right)}{\left( C(1)-C_{plateau}\right)}
\label{Cp}
\end{equation}

\noindent where $C'(t)=0$ characterizes the equilibrium nematic state (either stable or
metastable) . In Fig.\ref{corre5} we see the typical behavior of $C'(t)$ for different 
temperatures  from above $T_2$ down to temperatures well below $T_1$. Also $C'(t)$ exhibit 
different behaviors above and below $T_2$. For temperatures above $T_2$ the curves are 
very well fitted (see Fig.\ref{corre5}) by a modified stretched exponential function of 
the type

\begin{equation}
f(t) = A \; t^\omega e^{-\left(t/\tau \right)^\gamma}
\label{fit1}
\end{equation}

\begin{figure}
\begin{center}
\includegraphics[scale=0.35,angle=-90]{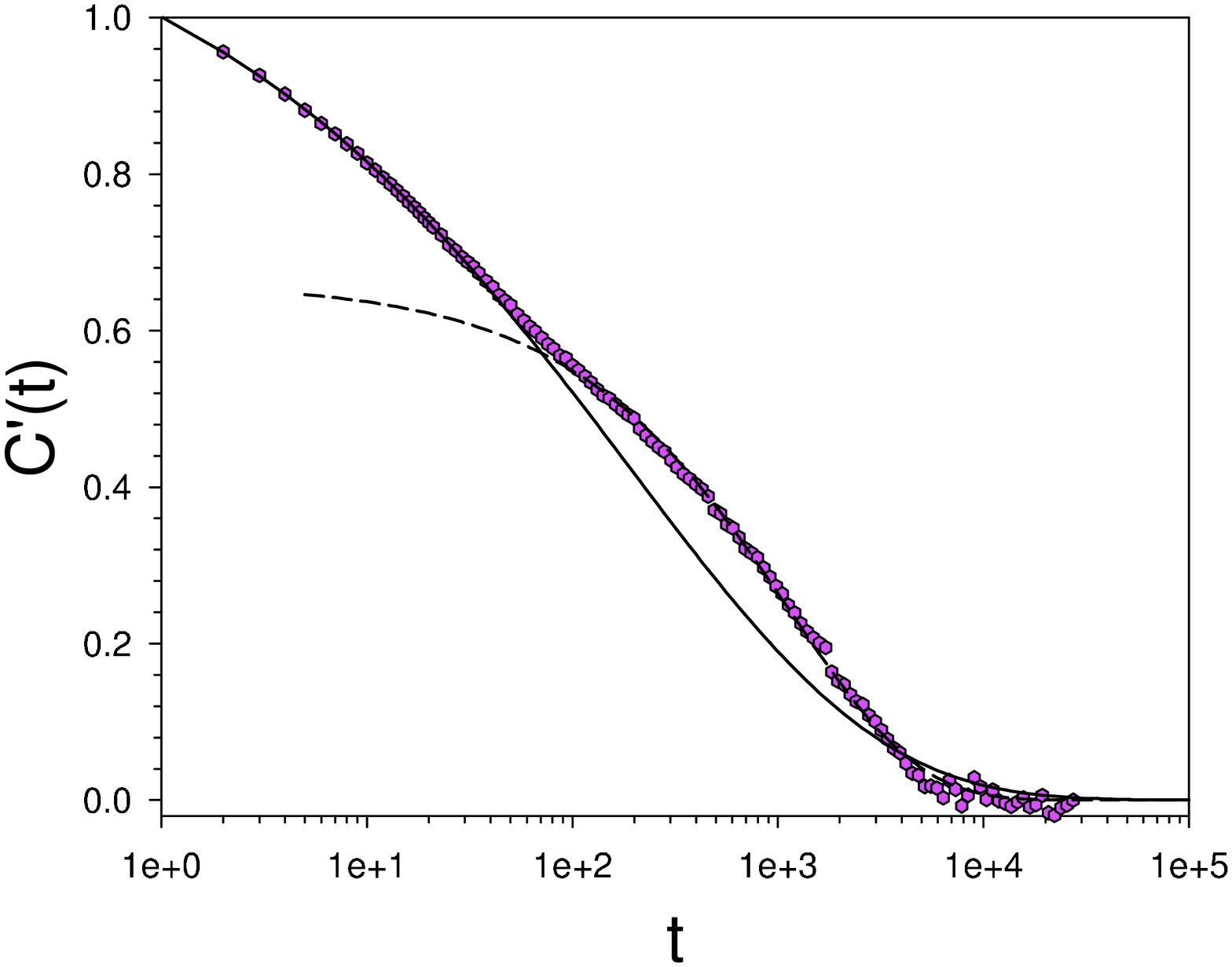}
\caption{\label{corre6} Normalized correlation function Eq.(\ref{Cp}) as a function of $t$ for $\delta=2$, after a quasi-static cooling at $T=0.78$. The continuous and dashed lines correspond to  fittings using Eqs.(\ref{fit1}) and (\ref{fit2}) respectively.}
\end{center}
\end{figure}

\noindent This fact is peculiar of the tetragonal liquid state: it does not exhibit exponential decay, even at relatively high temperatures. This behavior is observed also for $\delta=1$.  For temperatures below $T_1$, $C'(t)$ exhibit two different regimes, as can be appreciated more clearly in Fig.\ref{corre6}. At short times $C'(t)$ is again well fitted by a function of the type Eq.(\ref{fit1}). For times longer that some crossover time, $C'(t)$ enters into a pure stretched exponential regime, i.e., the best fit is obtained by a function of the type

\begin{equation}
f(t) = A \; e^{-\left(t/\tau \right)^\gamma}
\label{fit2}
\end{equation}

\begin{figure}[t]
\begin{center}
\includegraphics[scale=0.4,angle=-90]{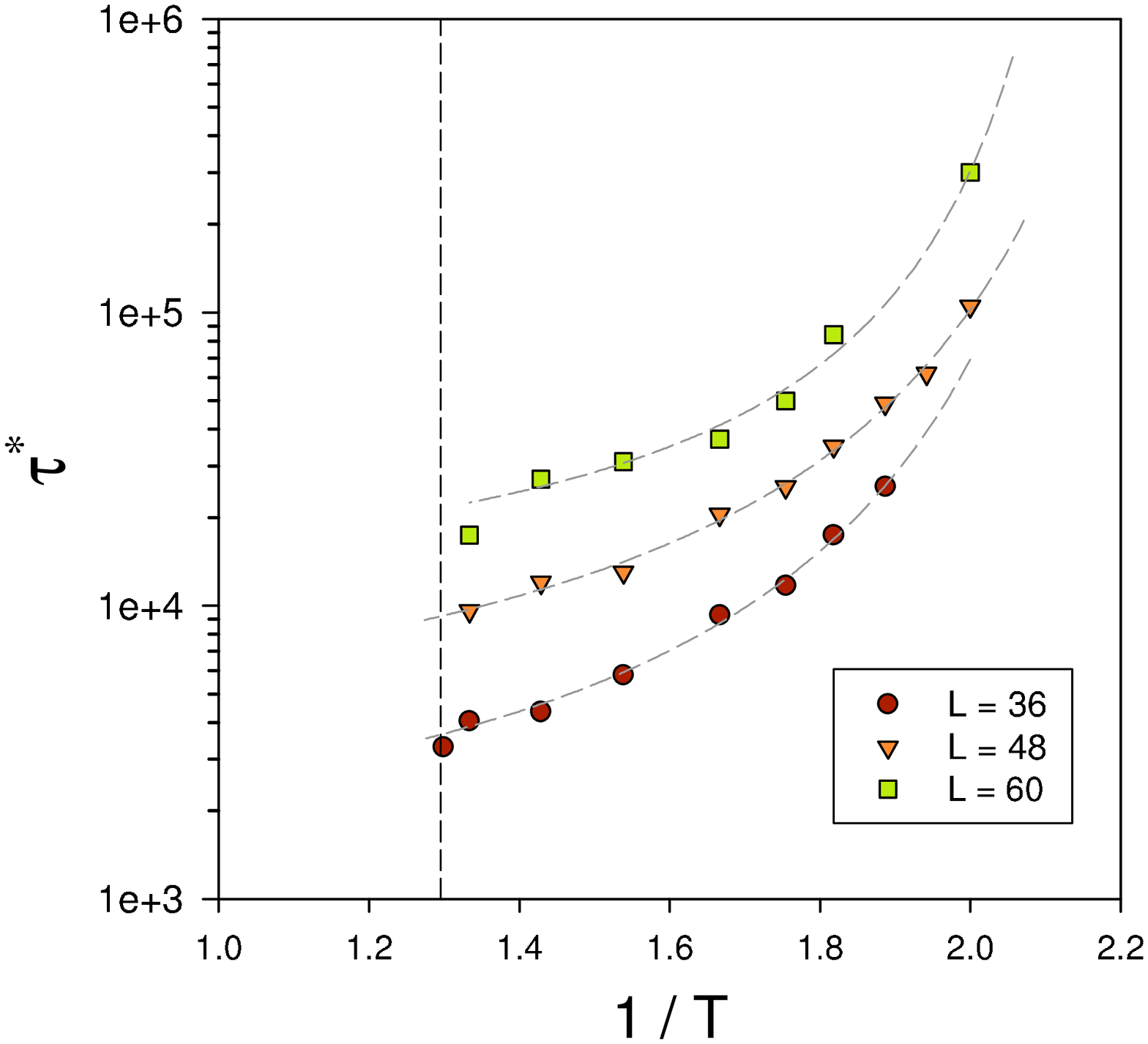}
\caption{\label{corretime} (Color online) Relaxation time $\tau^*$ vs. $1/T$ for
 $\delta=2$ and different system sizes.
The dashed lines correspond to fittings using VFT form. The vertical line corresponds to $1/T_1$}
\end{center}
\end{figure}

Finally, we analyzed the temperature dependence of the relaxation time $\tau$. Since there is
too much uncertainty in fitting a curve with three parameters, instead of considering $\tau$
we defined another correlation time $\tau^*$ as the time at which $C'(\tau^*)=0.1$.
In Fig.\ref{corretime} we  see an Arrhenius plot of $\tau^*$ for different system sizes. The
dashed lines in the figure correspond to fits using a Vogel-Fulcher-Tamman (VFT) form

\begin{equation}
\tau^* = \tau_0 \exp{\left( \frac{A}{T-T_0}\right)}
\label{VFT}
\end{equation}

\noindent  which is usually associated with the behavior of relaxation times in a
{\it fragile glass-former}, according to Angell's classification~\cite{AnNgMcMcMa2000}.
The fitting values of $T_0$ for the different system
sizes are shown in Table \ref{tabla2}. We also show the fragility parameter $K \equiv T_0/A$,
which show very high values around $K \approx 1$. It is worth mentioning that the data of Fig.\ref{corretime} can also be fitted with a power law of the type $\tau^*=A\, |T-T_0|^{-b}$. However, the best fitting is obtained with the VFT form Eq.(\ref{VFT}).

\begin{table}
\begin{center}
\begin{tabular}{|c|c|c|}
\hline
$L$ & $T_0$ & K \\
\hline
 $36$ & $0.41 \pm 0.025$ & $1.20\pm 0.4$\\
\hline
 $48$ & $0.40 \pm 0.015$ & $1.25\pm 0.3$\\
\hline
 $60$ & $0.44 \pm 0.075$ & $2.10 \pm 0.8$\\
\hline
\end{tabular}
\end{center}
\caption{Fitting values of $T_0$ and the fragility parameter $K\equiv
  T_0/A$ for the relaxation time
$\tau^*$ as a function of the supercooling temperature using a VFT form.}
\label{tabla2}
\end{table}

\begin{figure}[t]
\begin{center}
\includegraphics[scale=0.35,angle=-90]{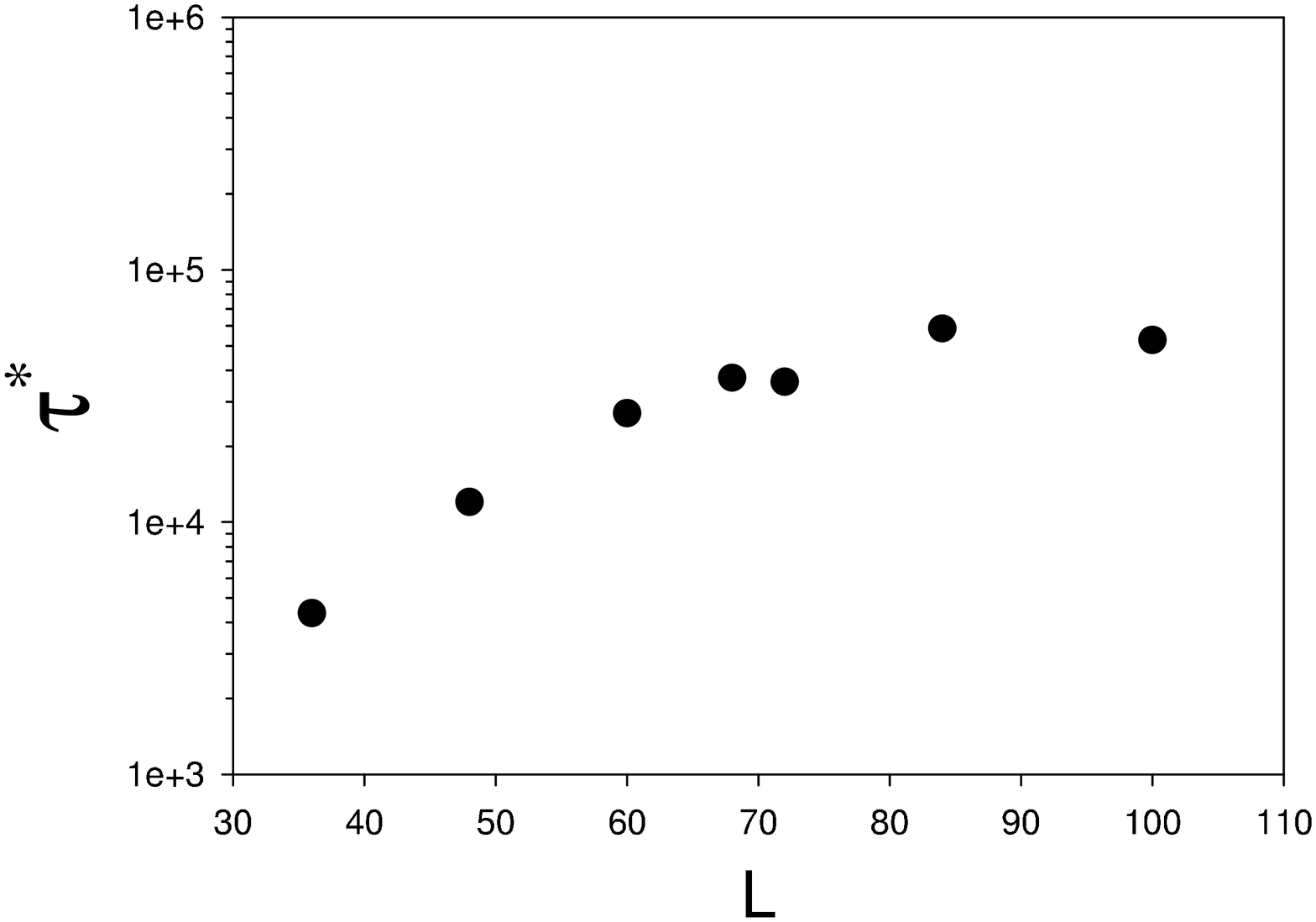}
\caption{\label{corretimeL} Relaxation time $\tau^*$ vs. $L$ for  $\delta=2$ and $T=0.7$.}
\end{center}
\end{figure}

Notice the strong dependence of $\tau^*$ with $L$ for the range of
sizes here considered. In
Fig.\ref{corretimeL} we show $\tau^*$ as a function of $L$ for a fixed
temperature in the metastable
region. We see that $\tau^*$ becomes independent of $L$ for sizes
larger than $L \sim 80$,
which coincides with the crossover value observed for the plateau,
suggesting the
same origin for both finite size effects.

\section{Discussion and conclusions}

The main result of this work is the observation of
interplay between nucleation and coarsening which
appears to be a novel feature. In other words, coarsening happens
usually as a competition between domains of different {\em stable}
phases. In the present case both direct
coarsening of the stripe phase and nucleation of the same phase on a
nematic background can happen with a
probability that depends on the final temperature. Nucleation processes are associated
with a strong metastability of the nematic phase. If the final temperature $T_f < T_1$
is not too
far below the stripe-nematic transition temperature the probability of nucleation
is high: during the quench the system initially relaxes to the nematic phase, which is
metastable below $T_1$ until finally it decays through nucleation to the equilibrium
stripe phase. For final temperatures deep in the stripe phase simple coarsening processes
between stripe states with different orientation have larger probability of occurrence
than nucleation. Although no detailed theory is
available for predicting this behavior in the whole temperature range, classical
arguments assuming homogeneous nucleation, coarsening with a non-conserved order
parameter and finite size scaling give a reasonably good interpretation of all the
observed phenomenology.
The origin of this interplay may rely in the fact that
the stable and the metastable phases share the same orientational
symmetry, so we can have domains with stripes (or pieces of
stripes) that can be oriented along the same directions of the square
lattice substrate in both phases.

Quasi-equilibrium autocorrelations in the stable and metastable nematic regions show
a plateau whose nature is not completely clear. For temperatures where the nematic
phase is only metastable, this plateau appears in a finite time window until a final
decay for times $t \geq \tau_{nucl}$. For temperatures in the equilibrium nematic phase
the behavior is more subtle. Although in a first approximation one could consider it
to be a distinct
characteristic of the nematic phase, in which a fraction of the spins remain practically
frozen, we got indications that for the larger sizes the height of the
plateau begins to diminish, signalling a possible finite size origin. One
possibility is that the fraction of frozen spins in the nematic phase diminish with
system size, in which case the plateau will eventually disappear for sizes large
enough. More work is needed to clarify the behavior of equilibrium relaxations in
large samples.

The present results also confirm the existence of the intermediate nematic phase, as
observed in equilibrium simulations in \cite{CaMiStTa2006,PiCa2007}; all the dynamical 
observations are in agreement with the reported equilibrium results. In fact, a
nematic phase with quasi-long-range order is expected to be generic in isotropic
systems with competing interactions~\cite{BaSt2007}. 
Recent analytic results on a continous version of the
present system show the presence of an isotropic-nematic transition which is in the
Kosterlitz-Thouless universality class, while the stripe phase with true positional
long range order turns out to be unstable for continous and isotropic
systems in $d=2$~\cite{BaSt2007}
in the thermodynamic limit, at variance with the discrete model studied here. 
Nevertheless, it is important to note that, although truly unstable in the thermodynamic
limit, stripe domains grow already in the paramagnetic phase with decreasing temperature
~\cite{MuSt2007}
and may be relevant for observations in finte size systems, like in simulations for example.
In the present case, the stability of the low temperature stripe phase may be due to the
symmetry breaking effect of the lattice and the sharp nature of the Ising domain walls. 
A unified interpretation of the nature of the
transitions in the discrete and continous models is still lacking.

A final important point concerns the strong increase observed in the stationary relaxation time 
$\tau^*$ in the metastable nematic phase. Although a spinodal instability cannot be excluded, 
the stretched exponential relaxation and the super Arrhenius behavior of $\tau^*(T)$ point
to an increasing cooperative and glassy behaviour of the dynamics. Indeed, a very similar 
behavior in a related model, namely the Coulomb frustrated model in 3d, has been interpreted 
as an evidence of fragile glass forming phenomenology\cite{GrTaVi2002}. Recent 
experimental results on Fe on Cu(100) ultrathin films\cite{PoVaPe2006} have suggested the 
possibility of a stripe-liquid to stripe-glass transition in the vicinity of the stripe 
ordering temperature. Stripe-glassy behaviour can be interpreted within the Frustration
Limited Domain Theory~\cite{TaKiNuVi2005} or the Random First Order Transition
~\cite{ScWo2000,WeScWo2001} scenarios. We 
have not attempted a quantitative comparison with either theory, which remains as subject
of future work. 

\section{Acknowledgments}

 This work was partially supported by grants from CONICET,
FONCyT grant PICT-2005 33305 , SeCyT-Universidad Nacional de C\'ordoba (Argentina), 
CNPq and CAPES (Brazil), and ICTP grant NET-61 (Italy).


\begin{thebibliography}{29}
\expandafter\ifx\csname natexlab\endcsname\relax\def\natexlab#1{#1}\fi
\expandafter\ifx\csname bibnamefont\endcsname\relax
  \def\bibnamefont#1{#1}\fi
\expandafter\ifx\csname bibfnamefont\endcsname\relax
  \def\bibfnamefont#1{#1}\fi
\expandafter\ifx\csname citenamefont\endcsname\relax
  \def\citenamefont#1{#1}\fi
\expandafter\ifx\csname url\endcsname\relax
  \def\url#1{\texttt{#1}}\fi
\expandafter\ifx\csname urlprefix\endcsname\relax\def\urlprefix{URL }\fi
\providecommand{\bibinfo}[2]{#2}
\providecommand{\eprint}[2][]{\url{#2}}

\bibitem[{\citenamefont{{Hubert} and {Schafer}}(1998)}]{HuSc1998}
\bibinfo{author}{\bibfnamefont{A.}~\bibnamefont{{Hubert}}} \bibnamefont{and}
  \bibinfo{author}{\bibfnamefont{R.}~\bibnamefont{{Schafer}}},
  \emph{\bibinfo{title}{Magnetic Domains}}
  (\bibinfo{publisher}{Springer-Verlag}, \bibinfo{address}{Berlin},
  \bibinfo{year}{1998}).

\bibitem[{\citenamefont{Politi}(1998)}]{Po1998}
\bibinfo{author}{\bibfnamefont{P.}~\bibnamefont{Politi}},
  \bibinfo{journal}{Comments Cond. Matter Phys.} \textbf{\bibinfo{volume}{18}},
  \bibinfo{pages}{191} (\bibinfo{year}{1998}).

\bibitem[{\citenamefont{Giuliani et~al.}(2006)\citenamefont{Giuliani, Lebowitz,
  and Lieb}}]{GiLeLi2006}
\bibinfo{author}{\bibfnamefont{A.}~\bibnamefont{Giuliani}},
  \bibinfo{author}{\bibfnamefont{J.~L.} \bibnamefont{Lebowitz}},
  \bibnamefont{and} \bibinfo{author}{\bibfnamefont{E.~H.} \bibnamefont{Lieb}},
  \bibinfo{journal}{Physical Review B} \textbf{\bibinfo{volume}{74}},
  \bibinfo{pages}{064420} (\bibinfo{year}{2006}).

\bibitem[{\citenamefont{Allenspach et~al.}(1990)\citenamefont{Allenspach,
  Stamponi, and Bischof}}]{AlStBi1990}
\bibinfo{author}{\bibfnamefont{R.}~\bibnamefont{Allenspach}},
  \bibinfo{author}{\bibfnamefont{M.}~\bibnamefont{Stamponi}}, \bibnamefont{and}
  \bibinfo{author}{\bibfnamefont{A.}~\bibnamefont{Bischof}},
  \bibinfo{journal}{Phys. Rev. Lett.} \textbf{\bibinfo{volume}{65}},
  \bibinfo{pages}{3344} (\bibinfo{year}{1990}).

\bibitem[{\citenamefont{Vaterlaus et~al.}(2000)\citenamefont{Vaterlaus, Stamm,
  Maier, Pini, Politi, and Pescia}}]{VaStMaPiPoPe2000}
\bibinfo{author}{\bibfnamefont{A.}~\bibnamefont{Vaterlaus}},
  \bibinfo{author}{\bibfnamefont{C.}~\bibnamefont{Stamm}},
  \bibinfo{author}{\bibfnamefont{U.}~\bibnamefont{Maier}},
  \bibinfo{author}{\bibfnamefont{M.~G.} \bibnamefont{Pini}},
  \bibinfo{author}{\bibfnamefont{P.}~\bibnamefont{Politi}}, \bibnamefont{and}
  \bibinfo{author}{\bibfnamefont{D.}~\bibnamefont{Pescia}},
  \bibinfo{journal}{Phys. Rev. Lett.} \textbf{\bibinfo{volume}{84}},
  \bibinfo{pages}{2247} (\bibinfo{year}{2000}).

\bibitem[{\citenamefont{Wu et~al.}(2004)\citenamefont{Wu, Won, Scholl, Doran,
  Zhao, Jin, and Qiu}}]{WuWoSc2004}
\bibinfo{author}{\bibfnamefont{Y.}~\bibnamefont{Wu}},
  \bibinfo{author}{\bibfnamefont{C.}~\bibnamefont{Won}},
  \bibinfo{author}{\bibfnamefont{A.}~\bibnamefont{Scholl}},
  \bibinfo{author}{\bibfnamefont{A.}~\bibnamefont{Doran}},
  \bibinfo{author}{\bibfnamefont{H.}~\bibnamefont{Zhao}},
  \bibinfo{author}{\bibfnamefont{X.}~\bibnamefont{Jin}}, \bibnamefont{and}
  \bibinfo{author}{\bibfnamefont{Z.}~\bibnamefont{Qiu}},
  \bibinfo{journal}{Physical Review Letters} \textbf{\bibinfo{volume}{93}},
  \bibinfo{pages}{117205} (\bibinfo{year}{2004}).

\bibitem[{\citenamefont{Won et~al.}(2005)\citenamefont{Won, Wu, Choi, Kim,
  Scholl, Doran, Owens, Wu, Jin, Zhao et~al.}}]{WoWuCh2005}
\bibinfo{author}{\bibfnamefont{C.}~\bibnamefont{Won}},
  \bibinfo{author}{\bibfnamefont{Y.}~\bibnamefont{Wu}},
  \bibinfo{author}{\bibfnamefont{J.}~\bibnamefont{Choi}},
  \bibinfo{author}{\bibfnamefont{W.}~\bibnamefont{Kim}},
  \bibinfo{author}{\bibfnamefont{A.}~\bibnamefont{Scholl}},
  \bibinfo{author}{\bibfnamefont{A.}~\bibnamefont{Doran}},
  \bibinfo{author}{\bibfnamefont{T.}~\bibnamefont{Owens}},
  \bibinfo{author}{\bibfnamefont{J.}~\bibnamefont{Wu}},
  \bibinfo{author}{\bibfnamefont{X.}~\bibnamefont{Jin}},
  \bibinfo{author}{\bibfnamefont{H.}~\bibnamefont{Zhao}}, \bibnamefont{et~al.},
  \bibinfo{journal}{Phys. Rev. B} \textbf{\bibinfo{volume}{71}},
  \bibinfo{pages}{224429} (\bibinfo{year}{2005}).

\bibitem[{\citenamefont{Portmann et~al.}(2006)\citenamefont{Portmann,
  Vaterlaus, and Pescia}}]{PoVaPe2006}
\bibinfo{author}{\bibfnamefont{O.}~\bibnamefont{Portmann}},
  \bibinfo{author}{\bibfnamefont{A.}~\bibnamefont{Vaterlaus}},
  \bibnamefont{and} \bibinfo{author}{\bibfnamefont{D.}~\bibnamefont{Pescia}},
  \bibinfo{journal}{Phys. Rev. Lett.} \textbf{\bibinfo{volume}{96}},
  \bibinfo{pages}{047212} (\bibinfo{year}{2006}).

\bibitem[{\citenamefont{Yafet and Gyorgy}(1988)}]{YaGy1988}
\bibinfo{author}{\bibfnamefont{Y.}~\bibnamefont{Yafet}} \bibnamefont{and}
  \bibinfo{author}{\bibfnamefont{E.~M.} \bibnamefont{Gyorgy}},
  \bibinfo{journal}{Phys. Rev. B} \textbf{\bibinfo{volume}{38}},
  \bibinfo{pages}{9145} (\bibinfo{year}{1988}).

\bibitem[{\citenamefont{Pescia and Pokrovsky}(1990)}]{PePo1990}
\bibinfo{author}{\bibfnamefont{D.}~\bibnamefont{Pescia}} \bibnamefont{and}
  \bibinfo{author}{\bibfnamefont{V.~L.} \bibnamefont{Pokrovsky}},
  \bibinfo{journal}{Phys. Rev. Lett.} \textbf{\bibinfo{volume}{65}},
  \bibinfo{pages}{2599} (\bibinfo{year}{1990}).

\bibitem[{\citenamefont{Abanov et~al.}(1995)\citenamefont{Abanov, Kalatsky,
  Pokrovsky, and Saslow}}]{AbKaPoSa1995}
\bibinfo{author}{\bibfnamefont{A.}~\bibnamefont{Abanov}},
  \bibinfo{author}{\bibfnamefont{V.}~\bibnamefont{Kalatsky}},
  \bibinfo{author}{\bibfnamefont{V.~L.} \bibnamefont{Pokrovsky}},
  \bibnamefont{and} \bibinfo{author}{\bibfnamefont{W.~M.}
  \bibnamefont{Saslow}}, \bibinfo{journal}{Phys. Rev. B}
  \textbf{\bibinfo{volume}{51}}, \bibinfo{pages}{1023} (\bibinfo{year}{1995}).

\bibitem[{\citenamefont{MacIsaac et~al.}(1995)\citenamefont{MacIsaac,
  Whitehead, Robinson, and De'Bell}}]{MaWhRoDe1995}
\bibinfo{author}{\bibfnamefont{A.~B.} \bibnamefont{MacIsaac}},
  \bibinfo{author}{\bibfnamefont{J.~P.} \bibnamefont{Whitehead}},
  \bibinfo{author}{\bibfnamefont{M.~C.} \bibnamefont{Robinson}},
  \bibnamefont{and} \bibinfo{author}{\bibfnamefont{K.}~\bibnamefont{De'Bell}},
  \bibinfo{journal}{Phys. Rev. B} \textbf{\bibinfo{volume}{51}},
  \bibinfo{pages}{16033} (\bibinfo{year}{1995}).

\bibitem[{\citenamefont{Cannas et~al.}(2004)\citenamefont{Cannas, Stariolo, and
  Tamarit}}]{CaStTa2004}
\bibinfo{author}{\bibfnamefont{S.~A.} \bibnamefont{Cannas}},
  \bibinfo{author}{\bibfnamefont{D.~A.} \bibnamefont{Stariolo}},
  \bibnamefont{and} \bibinfo{author}{\bibfnamefont{F.~A.}
  \bibnamefont{Tamarit}}, \bibinfo{journal}{Phys. Rev. B}
  \textbf{\bibinfo{volume}{69}}, \bibinfo{pages}{092409}
  (\bibinfo{year}{2004}).

\bibitem[{\citenamefont{Cannas et~al.}(2006)\citenamefont{Cannas, Michelon,
  Stariolo, and Tamarit}}]{CaMiStTa2006}
\bibinfo{author}{\bibfnamefont{S.~A.} \bibnamefont{Cannas}},
  \bibinfo{author}{\bibfnamefont{M.}~\bibnamefont{Michelon}},
  \bibinfo{author}{\bibfnamefont{D.~A.} \bibnamefont{Stariolo}},
  \bibnamefont{and} \bibinfo{author}{\bibfnamefont{F.~A.}
  \bibnamefont{Tamarit}}, \bibinfo{journal}{Phys. Rev. B}
  \textbf{\bibinfo{volume}{73}}, \bibinfo{pages}{184425}
  (\bibinfo{year}{2006}),
  \urlprefix\url{http://link.aps.org/abstract/PRB/v73/e184425}.

\bibitem[{\citenamefont{Rastelli et~al.}(2006)\citenamefont{Rastelli, Regina,
  and Tassi}}]{RaReTa2006}
\bibinfo{author}{\bibfnamefont{E.}~\bibnamefont{Rastelli}},
  \bibinfo{author}{\bibfnamefont{S.}~\bibnamefont{Regina}}, \bibnamefont{and}
  \bibinfo{author}{\bibfnamefont{A.}~\bibnamefont{Tassi}},
  \bibinfo{journal}{Phys. Rev. B} \textbf{\bibinfo{volume}{73}},
  \bibinfo{pages}{144418} (\bibinfo{year}{2006}).

\bibitem[{\citenamefont{Pighin and Cannas}(2007)}]{PiCa2007}
\bibinfo{author}{\bibfnamefont{S.~A.} \bibnamefont{Pighin}} \bibnamefont{and}
  \bibinfo{author}{\bibfnamefont{S.~A.} \bibnamefont{Cannas}},
  \bibinfo{journal}{Physical Review B (Condensed Matter and Materials Physics)}
  \textbf{\bibinfo{volume}{75}}, \bibinfo{eid}{224433}
  (pages~\bibinfo{numpages}{9}) (\bibinfo{year}{2007}),
  \urlprefix\url{http://link.aps.org/abstract/PRB/v75/e224433}.

\bibitem[{\citenamefont{Carubelli et~al.}(2008)\citenamefont{Carubelli,
  Billoni, Pighin, Cannas, Stariolo, and Tamarit}}]{CaBiPiCaStTa2008}
\bibinfo{author}{\bibfnamefont{M.}~\bibnamefont{Carubelli}},
  \bibinfo{author}{\bibfnamefont{O.~V.} \bibnamefont{Billoni}},
  \bibinfo{author}{\bibfnamefont{S.}~\bibnamefont{Pighin}},
  \bibinfo{author}{\bibfnamefont{S.~A.} \bibnamefont{Cannas}},
  \bibinfo{author}{\bibfnamefont{D.~A.} \bibnamefont{Stariolo}},
  \bibnamefont{and} \bibinfo{author}{\bibfnamefont{F.~A.}
  \bibnamefont{Tamarit}}, \bibinfo{journal}{Phys. Rev. B}
  \textbf{\bibinfo{volume}{77}}, \bibinfo{pages}{134417}
  (\bibinfo{year}{2008}).

\bibitem[{\citenamefont{Sampaio et~al.}(1996)\citenamefont{Sampaio,
  de~Albuquerque, and de~Menezes}}]{SaAlMe1996}
\bibinfo{author}{\bibfnamefont{L.~C.} \bibnamefont{Sampaio}},
  \bibinfo{author}{\bibfnamefont{M.~P.} \bibnamefont{de~Albuquerque}},
  \bibnamefont{and} \bibinfo{author}{\bibfnamefont{F.~S.}
  \bibnamefont{de~Menezes}}, \bibinfo{journal}{Phys. Rev. B}
  \textbf{\bibinfo{volume}{54}}, \bibinfo{pages}{6465} (\bibinfo{year}{1996}).

\bibitem[{\citenamefont{Toloza et~al.}(1998)\citenamefont{Toloza, Tamarit, and
  Cannas}}]{ToTaCa1998}
\bibinfo{author}{\bibfnamefont{J.~H.} \bibnamefont{Toloza}},
  \bibinfo{author}{\bibfnamefont{F.~A.} \bibnamefont{Tamarit}},
  \bibnamefont{and} \bibinfo{author}{\bibfnamefont{S.~A.}
  \bibnamefont{Cannas}}, \bibinfo{journal}{Phys. Rev. B}
  \textbf{\bibinfo{volume}{58}}, \bibinfo{pages}{R8885} (\bibinfo{year}{1998}).

\bibitem[{\citenamefont{Stariolo and Cannas}(1999)}]{StCa1999}
\bibinfo{author}{\bibfnamefont{D.~A.} \bibnamefont{Stariolo}} \bibnamefont{and}
  \bibinfo{author}{\bibfnamefont{S.~A.} \bibnamefont{Cannas}},
  \bibinfo{journal}{Phys. Rev. B} \textbf{\bibinfo{volume}{60}},
  \bibinfo{pages}{3013} (\bibinfo{year}{1999}).

\bibitem[{\citenamefont{Gleiser et~al.}(2003)\citenamefont{Gleiser, Tamarit,
  Cannas, and Montemurro}}]{GlTaCaMo2003}
\bibinfo{author}{\bibfnamefont{P.~M.} \bibnamefont{Gleiser}},
  \bibinfo{author}{\bibfnamefont{F.~A.} \bibnamefont{Tamarit}},
  \bibinfo{author}{\bibfnamefont{S.~A.} \bibnamefont{Cannas}},
  \bibnamefont{and} \bibinfo{author}{\bibfnamefont{M.~A.}
  \bibnamefont{Montemurro}}, \bibinfo{journal}{Phys. Rev. B}
  \textbf{\bibinfo{volume}{68}}, \bibinfo{pages}{134401}
  (\bibinfo{year}{2003}).

\bibitem[{\citenamefont{Lee and Kosterlitz}(1991)}]{LeKo1991}
\bibinfo{author}{\bibfnamefont{J.}~\bibnamefont{Lee}} \bibnamefont{and}
  \bibinfo{author}{\bibfnamefont{J.~M.} \bibnamefont{Kosterlitz}},
  \bibinfo{journal}{Phys. Rev. B} \textbf{\bibinfo{volume}{43}},
  \bibinfo{pages}{3265} (\bibinfo{year}{1991}).

\bibitem[{\citenamefont{Angell et~al.}(2000)\citenamefont{Angell, Ngai, Kenna,
  Millan, and Martin}}]{AnNgMcMcMa2000}
\bibinfo{author}{\bibfnamefont{C.~A.} \bibnamefont{Angell}},
  \bibinfo{author}{\bibfnamefont{K.~L.} \bibnamefont{Ngai}},
  \bibinfo{author}{\bibfnamefont{G.~B.~M.} \bibnamefont{Kenna}},
  \bibinfo{author}{\bibfnamefont{P.~F.~M.} \bibnamefont{Millan}},
  \bibnamefont{and} \bibinfo{author}{\bibfnamefont{S.~W.}
  \bibnamefont{Martin}}, \bibinfo{journal}{Journal of Applied Physics}
  \textbf{\bibinfo{volume}{88}}, \bibinfo{pages}{3113} (\bibinfo{year}{2000}).

\bibitem[{\citenamefont{Barci and Stariolo}(2007)}]{BaSt2007}
\bibinfo{author}{\bibfnamefont{D.~G.} \bibnamefont{Barci}} \bibnamefont{and}
  \bibinfo{author}{\bibfnamefont{D.~A.} \bibnamefont{Stariolo}},
  \bibinfo{journal}{Physical Review Letters} \textbf{\bibinfo{volume}{98}},
  \bibinfo{eid}{200604} (pages~\bibinfo{numpages}{4}) (\bibinfo{year}{2007}),
  \urlprefix\url{http://link.aps.org/abstract/PRL/v98/e200604}.

\bibitem[{\citenamefont{Mulet and Stariolo}(2007)}]{MuSt2007}
\bibinfo{author}{\bibfnamefont{R.}~\bibnamefont{Mulet}} \bibnamefont{and}
  \bibinfo{author}{\bibfnamefont{D.~A.} \bibnamefont{Stariolo}},
  \bibinfo{journal}{Physical Review B (Condensed Matter and Materials Physics)}
  \textbf{\bibinfo{volume}{75}}, \bibinfo{eid}{064108}
  (pages~\bibinfo{numpages}{12}) (\bibinfo{year}{2007}),
  \urlprefix\url{http://link.aps.org/abstract/PRB/v75/e064108}.

\bibitem[{\citenamefont{Grousson et~al.}(2002)\citenamefont{Grousson, Tarjus,
  and Viot}}]{GrTaVi2002}
\bibinfo{author}{\bibfnamefont{M.}~\bibnamefont{Grousson}},
  \bibinfo{author}{\bibfnamefont{G.}~\bibnamefont{Tarjus}}, \bibnamefont{and}
  \bibinfo{author}{\bibfnamefont{P.}~\bibnamefont{Viot}},
  \bibinfo{journal}{Phys. Rev. E} \textbf{\bibinfo{volume}{65}},
  \bibinfo{pages}{065103} (\bibinfo{year}{2002}).

\bibitem[{\citenamefont{Tarjus et~al.}(2005)\citenamefont{Tarjus, Kivelson,
  Nussinov, and Viot}}]{TaKiNuVi2005}
\bibinfo{author}{\bibfnamefont{G.}~\bibnamefont{Tarjus}},
  \bibinfo{author}{\bibfnamefont{S.~A.} \bibnamefont{Kivelson}},
  \bibinfo{author}{\bibfnamefont{Z.}~\bibnamefont{Nussinov}}, \bibnamefont{and}
  \bibinfo{author}{\bibfnamefont{P.}~\bibnamefont{Viot}},
  \bibinfo{journal}{Journal of Physics: Condensed Matter}
  \textbf{\bibinfo{volume}{17}}, \bibinfo{pages}{R1143} (\bibinfo{year}{2005}).

\bibitem[{\citenamefont{Schmalian and Wolynes}(2000)}]{ScWo2000}
\bibinfo{author}{\bibfnamefont{J.}~\bibnamefont{Schmalian}} \bibnamefont{and}
  \bibinfo{author}{\bibfnamefont{P.~G.} \bibnamefont{Wolynes}},
  \bibinfo{journal}{Phys. Rev. Lett.} \textbf{\bibinfo{volume}{85}},
  \bibinfo{pages}{836} (\bibinfo{year}{2000}).

\bibitem[{\citenamefont{Westfahl~Jr et~al.}(2001)\citenamefont{Westfahl~Jr,
  Schmalian, and Wolynes}}]{WeScWo2001}
\bibinfo{author}{\bibfnamefont{H.}~\bibnamefont{Westfahl~Jr}},
  \bibinfo{author}{\bibfnamefont{J.}~\bibnamefont{Schmalian}},
  \bibnamefont{and} \bibinfo{author}{\bibfnamefont{P.~G.}
  \bibnamefont{Wolynes}}, \bibinfo{journal}{Phys. Rev. B}
  \textbf{\bibinfo{volume}{64}}, \bibinfo{pages}{174203}
  (\bibinfo{year}{2001}).

\end{thebibliography}

\end{document}